\documentclass[twocolumn,showpacs,preprintnumbers,amsmath,amssymb]{revtex4}

\usepackage{graphicx}
\usepackage{dcolumn}
\usepackage{bm}

\begin{document}


\title{Neutrino Nucleus Reactions based on New Shell Model 
Hamiltonians}

\author{Toshio Suzuki}
 \email{suzuki@chs.nihon-u.ac.jp}
\affiliation{Department of Physics, College of Humanities and
 Sciences, Nihon University\\     
Sakurajosui 3-25-40, Setagaya-ku, Tokyo 156-8550, Japan}
%
\author{Satoshi Chiba}
\affiliation{Advanced Science Research Center, Japan Atomic Energy Agency\\    
2-4 Shirakata-shirane, Tokai, Naka-gun, Ibaraki 319-1195, Japan}
\author{Takashi Yoshida}
\affiliation{National Astronomical Observatory of Japan, 
Mitaka, Tokyo 181-8588, 
Japan}
\author{Toshitaka Kajino}
\affiliation{National Astronomical Observatory of Japan, 
Mitaka, Tokyo 181-8588, 
Japan\\
Deaprtment of Astronomy, Graduate School of Science, University of Tokyo, 
Bunkyo-ku, Tokyo 113-0033,
Japan}
\author{Takaharu Otsuka}
\affiliation{Department of Physics, University of Tokyo, Hongo, 
Bunkyo-ku, Tokyo 113-0033, Japan\\
Center for Nuclear Study, University of Tokyo, Hongo, Bunkyo-ku,
Tokyo 113-0033, Japan\\
RIKEN, Hirosawa, Wako-shi, Saitama 351-0198, Japan}


\date{\today}

\begin{abstract}

A new shell model Hamiltonian for $p$-shell nuclei which properly
takes into account important roles of spin-isospin interactions
is used to obtain cross sections of neutrino-$^{12}$C reactions
induced by decay-at-rest (DAR) neutrinos as well as supernova
neutrinos. Branching ratios to various decay channels are  
calculated by the Hauser-Feshbach theory. 
Neutrino -$^{4}$He reactions are also investigated by using 
recent shell model Hamiltonians. The reaction cross sections
are found to be enhanced for both $^{12}$C and $^{4}$He compared
with previous calculations. 
As an interesting consequence of this,  
a possible enhancement of the
production yields of light elements, $^{7}$Li 
and $^{11}$B, during supernova
explosions is pointed out.    
 
\end{abstract}

\maketitle


\def\be{\begin{equation}}
\def\ee{\end{equation}}
\def\bea{\begin{eqnarray}}
\def\eea{\end{eqnarray}}
\def\br{\bf r}


    



\section{INTRODUCTION}

A recent progress in shell model calculations, which properly 
takes into account important roles of spin-isospin interactions,
is found to lead to significant improvements in magnetic 
properties of nuclei \cite{SFO03} as well as proper shell evolution,
that is, the change of magic numbers toward the drip-lines \cite{OUB,
OSFGA}. An important general role of the tensor interaction is
pointed out \cite{OSFGA}.   
The modified Hamiltonian can explain spin properties of the 
$p$-shell nuclei such as Gamow-Teller transitions better than 
conventional shell model Hamiltonians. In particular, agreements
between calculated and observed magnetic moments are found to be
systematically improved for the $p$-shell nuclei \cite{SFO03}. 
 
Here, we study new ingredients of these developments on 
neutrino-nucleus reactions on $^{12}$C, which are dominantly 
induced by Gamow-Teller and spin-dipole transitions. 
Charge-exchange and neutral current reactions induced by neutrinos
from pion 
decay-at rest (DAR) and supernova neutrinos are investigated, 
and comparisons are made with previous calculations \cite{VACSG,  
HT,WHHH}. We also study neutrino-nucleus recations on $^{4}$He, 
which are mainly induced by spin-dipole transitions. 
We discuss possible effects of our new neutrino-nucleus
reaction cross sections on nucleosynthesis process, in particular, 
on light element abundances during
supernova explosions.  

The paper is organized as follows. 
In section II, we discuss tensor components of our modified 
interaction. Neutrino-nucleus reaction cross sections on $^{12}$C
and $^{4}$He are obtained by using our new shell model Hamiltonian
in section III.  
Astrophysical implications are discussed in section IV, and 
a summary is given in section V.

\section{NEW SHELL MODEL HAMILTONIAN FOR $p$-SHELL NUCLEI}
 
We discuss some important ingredients of our new shell model
Hamiltonian for $p$-shell nuclei in ref. \cite{SFO03}, where
the spin-isospin flip interaction and the shell gap between the
$0p_{1/2}$ and $0p_{3/2}$ orbits are enhanced in comparison to
the Cohen-Kurath
Hamiltonian, (8-16)2BME \cite{CK}. We will refer to this 
Hamiltonian as SFO hereafter. 

\begin{figure*}[tbh]
\vspace{-30mm}
\hspace{-15mm}
\includegraphics[scale=0.53]{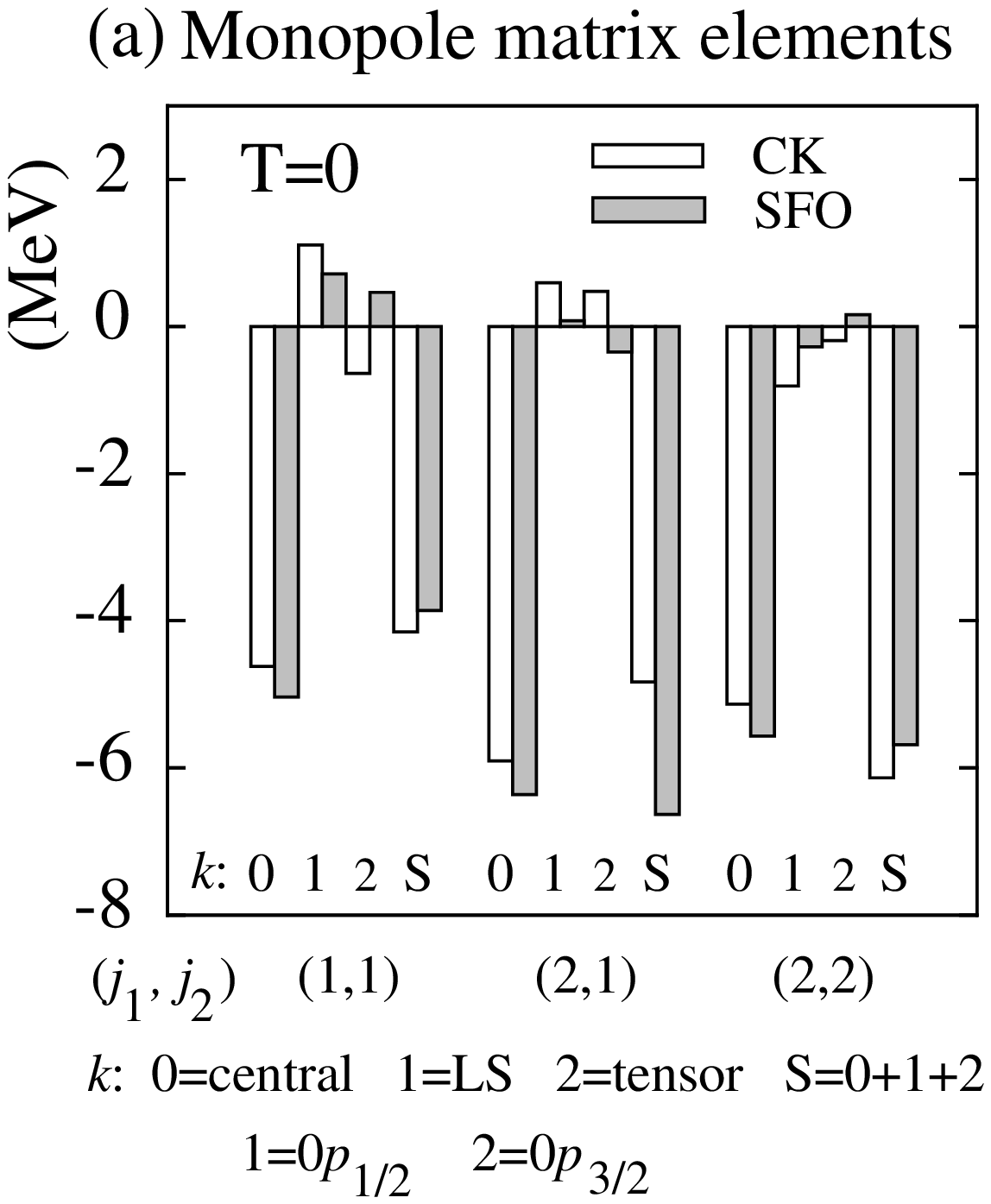}
\vspace{-30mm}
\hspace{-15mm}
\includegraphics[scale=0.53]{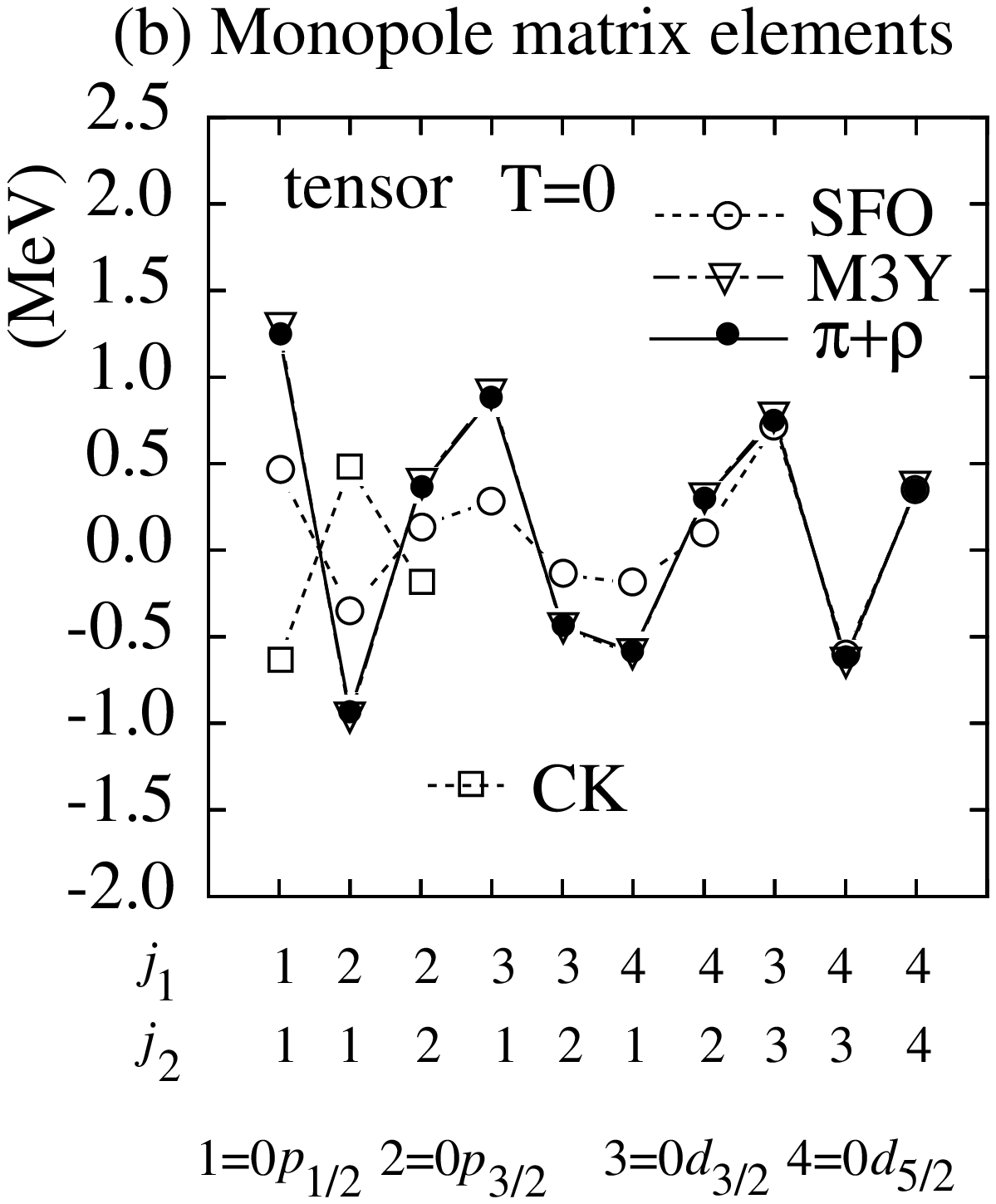}
\vspace{2.1cm}
\caption{(a) Monopole terms of the central ($k$=0), spin-orbit ($k$=1) 
and tensor ($k$=2) components of the SFO and Cohen-Kurath interactions. 
(b) Monopole terms of the tensor component of the SFO and Cohen-Kurath
interactions as well as the $\pi + \rho$ exchange potential and the
M3Y interaction.  
\label{fig:fig1}}
\end{figure*}

\begin{figure*}[tbh]
\vspace{-40mm}
\hspace{-15mm}
\includegraphics[scale=0.55]{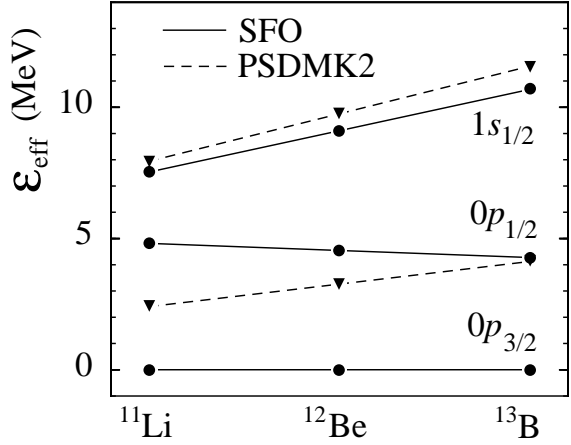}
\vspace{-1.2cm}
\caption{Effective neutron single particle energies relative to that
of the $0p_{3/2}$ orbit for N=8 isotones.
\label{fig:fig2}}
\end{figure*}

First, we show that the dominant effect of the enhancement of
the spin-flip two-body matrix elements in the isospin $T$=0 
channel, $\langle 0p_{3/2}0p_{1/2}: JT \mid V \mid 0p_{3/2}
0p_{1/2}: JT \rangle$ with $J$=1, 2, is the modification of
the tensor component of the interaction. Spin-tensor decomposition 
of the two-body effective interactions for a specific isospin 
channel can be done by expanding the matrix element with those
of the same orbital angular momenta \cite{Kir,Elliot}. 
Each matrix element is decomposed into the central ($k$=0), 
spin-orbit ($k$=1) and tensor ($k$=2) components, 
\begin{equation}
\langle ab: JT \mid V \mid cd: JT \rangle = \sum_{k}
\langle ab: JT \mid V_{k} \mid cd: JT \rangle
\end{equation}
with
\begin{eqnarray}
& &\langle ab: JT \mid V_{k} \mid cd: JT \rangle = (-1)^{J}(2k+1)
\sum_{LL'SS'} \nonumber\\
& &\langle ab\mid LSJ\rangle \langle cd\mid L'S'J\rangle
\left\{\begin{array}{ccc}
L & S & J\\
S' & L' & k
\end{array}\right\} 
\sum_{J'}(-1)^{J'}\nonumber\\
& &(2J'+1)
\left\{\begin{array}{ccc}
L & S & J'\\
S' & L' & k
\end{array}\right\} \sum_{j_{a}'j_{b}'j_{c}'j_{d}'}
\langle a'b'\mid LSJ'\rangle\nonumber\\ 
& & \langle c'd'\mid L'S'J'\rangle
\langle a'b': J'T \mid V \mid c'd': J'T\rangle 
\end{eqnarray}
where $a=\{n_a \ell_a j_a\}$, $a'=\{n_a \ell_a j_a'\}$, 
and   
$\langle ab\mid LSJ\rangle = 
\left\{\begin{array}{ccc}
\ell_{a} & 1/2 & j_{a}\\
\ell_{b} & 1/2 & j_{b}\\
L & S & J
\end{array}\right\}
\hat{j_a}\hat{j_b}\hat{L}\hat{S}$
with
$\hat{L}=2L+1$ etc. 

The monopole terms of the three components
\begin{equation}
V_{k}^{j_1 j_2,M} = \frac{\sum_{J} (2J+1)\langle j_1 j_2: JT \mid V_{k} 
\mid j_1 j_2: JT \rangle}{\sum_{J}(2J+1)}
\end{equation}
are shown in Fig. 1a for the $p$-shell matrix elements with $T$=0
for the SFO and the original Cohen-Kurath Hamiltonian. 
We find that the most important modification appears in the
tensor components where even the signs of the matrix elements
are changed. The central components of the monopole terms are
also increased. Note that the total $p$-shell matrix elements
are renormalized by a factor 0.93 \cite{SFO03}. 


The attractive (repulsive) nature of the tensor components 
of the monopole matrix elements with $j_1=j_{\rangle} = 0p_{3/2}$
and $j_2=j_{\langle} = 0p_{1/2}$ ($j_1=j_2=j_{\rangle} =0p_{3/2}$
or $j_1=j_2=j_{\langle} =0p_{1/2}$) is 
consistent with the general robust nature of the tensor
interaction \cite{OSFGA}. 
  
Although the magnitude of the tensor components of the monopole 
matrix elements of the SFO Hamiltonian is small compared with
that of the pion and rho-meson ($\pi + \rho$) exchange potential
with a radial cut off at 0.7 fm \cite{PIRHO} and the M3Y 
interaction \cite{M3Y}, their signs and the zigzag pattern of the 
monopole matrix element as a function of $j_{\rangle}-j_{\langle}$ 
and $j_{\rangle}-j_{\rangle}$ 
or $j_{\langle}-j_{\langle}$ are consistent as
shown in Fig. 1b. 
This zigzag structure with the proper signs in the tensor 
monopole terms is important and essential for the proper shell
evolution.    


Next, we show how these important characteristics affect the 
behavior of the effective single particle energies. 
Effective neutron single particle energies for $N$=8 isotones 
are shown in Fig. 2. The effective single particle energy is
the sum of the bare single particle energy for the $^{4}$He core
and monopole two-body matrix elements of the
proton-neutron (p-n) interaction summed over occupied proton
orbits outside the $^{4}$He core. Since the tensor interaction
is attractive between the proton ($\pi$) $0p_{3/2}$ orbit and
neutron ($\nu$) $0p_{1/2}$ orbit while it is repulsive between
the $\pi0p_{3/2}$ and $\nu0p_{3/2}$ orbits, the energy gap 
between the $\nu0p_{1/2}$ and $\nu0p_{3/2}$ orbits becomes 
larger as the proton number gets smaller, that is, for more 
neutron-rich isotones.  
The monopole terms of the central interaction are attractive 
both for $j_1=\pi0p_{3/2}$, $j_2=\nu0p_{1/2}$ and 
$j_1=\pi0p_{3/2}$, $j_2=\nu0p_{3/2}$, and their difference 
has the same sign as the tensor interaction but the magnitude
is smaller about by half. 
The monopole terms of the spin-orbit interaction work opposite 
to the tensor and central interactions. The proper shell
evolution is not obtained in case of the original Cohen-Kurath
interaction as the monopole terms of the tensor components 
have opposite signs compared to the SFO interaction, which
results in the reduction of the energy gap between the
$\nu0p_{1/2}$ and $\nu0p_{3/2}$ orbits in the neutron-rich 
side.   

We shall now go on to the question of to what extent such a
shell evolution is related to neutrino-nucleus reactions.

\section{NEUTRINO NUCLEUS REACTIONS}
\subsection{REACTIONS ON $^{12}$C INDUCED BY DAR NEUTRINOS}

We showed in ref.\cite{SFO03} that the magnetic properties of $p$-shell
nuclei are considerably improved, for example, in magnetic moments
and Gamow-Teller transitions. Here, we study another example of 
spin dependent transitions, namely neutrino-nucleus reactions, which 
are induced mainly by excitations of Gamow-Teller and spin-dipole
states. 


\begin{table*}
\caption{\label{tab:table1}
Cross sections for the exclusive reaction $^{12}$C ($\nu_{e}, e^{-}$)
$^{12}$N ($1_{g.s.}^{+}$) obtained for DAR neutrinos 
by shell model calculations. 
The bare $g_{A}$ is used except for the SFO Hamiltonian case with 
the value of $g_{A}^{eff}$ specified as $g_{A}^{eff}$ =0.95 $g_{A}$. 
Experimental values are taken from refs. \cite{LSND} and \cite{KARMEN}. 
The first error is statistical and the second is systematic.} 
\begin{tabular}{ c|c } \hline
Hamiltonian & cross section ($\times 10^{-42}$ cm$^{2}$) \\\hline  
SFO & 9.96  \\\hline
SFO ($g_{A}^{eff}$ =0.95 $g_{A}$) & 9.06 \\\hline
PSDMK2 & 8.48 \\\hline
WBT \cite{VACSG} & 8.42 \\\hline
Hayes-Towner \cite{HT} & 8.40 \\\hline
experiment (LSND \cite{LSND}) & 8.9$\pm$0.3$\pm$0.9 \\\hline
experiment (KARMEN \cite{KARMEN}) & 9.1$\pm$0.5$\pm$0.8 \\\hline
\end{tabular}
\end{table*}

\begin{table*}
\caption{\label{tab:table2}
Cross sections for the exclusive neutral current reaction on $^{12}$C 
leading to the 1$^{+}$ ($T$=1, 15.1 MeV) state induced by DAR neutrinos;
$\nu_{e}$'s and $\bar{\nu}_{\mu}$'s from $\mu^{+}$ decay as well as 
$\nu_{\mu}$'s from $\pi^{+}$ decay.  
Experimental values of the sum of $\nu_{e}$- and $\bar{\nu}_{\mu}$-induced 
reaction cross sections and that of $\nu_{\mu}$-induced reaction cross section
are taken from refs. \cite{KARMEN} and \cite{KARM2}, respectively.} 
\begin{tabular}{ c|c|c|c|c } \hline
Hamiltonian & \multicolumn{4}{c} 
{cross sections ($\times 10^{-42}$ cm$^{2}$)} \\
\cline{2-5} & \hfil
($\nu_{e}, \nu_{e}'$) & ($\bar{\nu}_{\mu}, \bar{\nu}_{\mu}'$) & 
($\nu_{e}, \nu_{e}'$) + ($\bar{\nu}_{\mu}, \bar{\nu}_{\mu}'$) & 
($\nu_{\mu}, \nu_{\mu}'$) \\\hline  
SFO ($g_{A}^{eff}$ =0.95 $g_{A}$) & 4.44 & 5.32 & 9.76 & 2.68 \\\hline
PSDMK2 & 3.75 & 4.52 & 8.27 & 2.26 \\\hline
experiment (KARMEN) & & & 10.4$\pm$1.0$\pm$0.9 \cite{KARMEN}& 
3.2$\pm$0.5$\pm$0.4 \cite{KARM2} \\\hline
\end{tabular}
\end{table*}

\begin{table*}
\caption{\label{tab:table3}
Cross sections for the reaction process $^{12}$C ($\nu_{e}, e^{-}$)
$^{12}$N$^{\ast}$ obtained for DAR neutrinos by shell model calculations. 
The bare $g_{A}$ is used unless specified. Experimental values are
taken from refs. \cite{LSND} and \cite{KARM3}.} 
\begin{tabular}{ c|c } \hline
Hamiltonian & cross section ($\times 10^{-42}$ cm$^{2}$) \\\hline
SFO & 8.35 \\\hline
SFO ($g_{A}^{eff}$ =0.70$g_{A}$) & 5.22 \\\hline
PSDMK2 & 7.14 \\\hline
PSDMK2 ($g_{A}^{eff}$ =0.75$g_{A}$) & 4.87 \\\hline
WBT \cite{VACSG} & 8.31 \\\hline
Hayes-Towner \cite{HT} & 3.80 \\\hline
experiment (LSND \cite{LSND}) & 4.3$\pm$0.4$\pm$0.6 \\\hline
experiment (KARMEN \cite{KARM3}) & 5.1$\pm$0.6$\pm$0.5 \\\hline
\end{tabular}
\end{table*}

\begin{figure*}[tbh]
\vspace{-35mm}
\hspace{-10mm}
\includegraphics[scale=0.57]{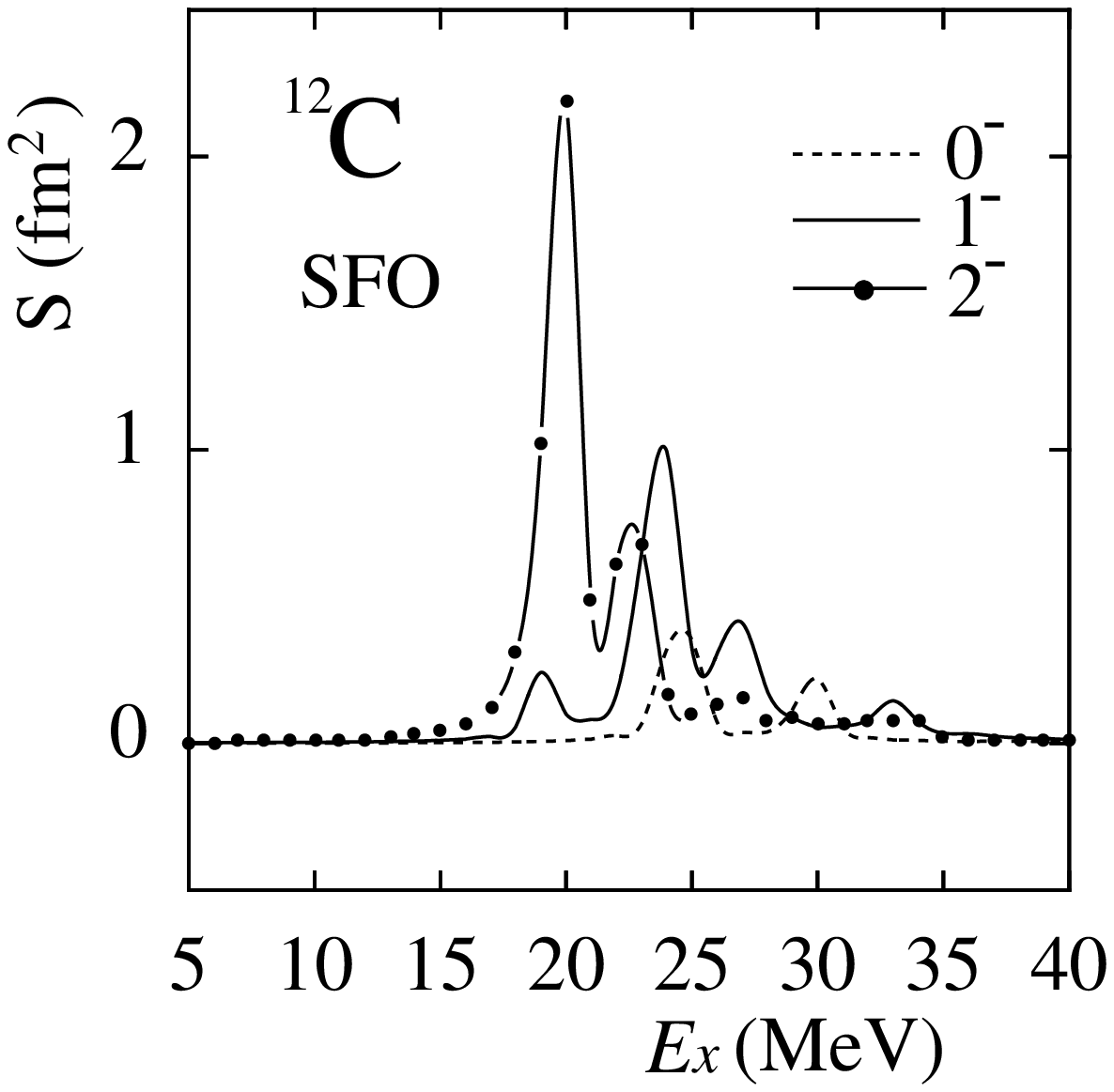}
\vspace{-35mm}
\hspace{-15mm}
\includegraphics[scale=0.57]{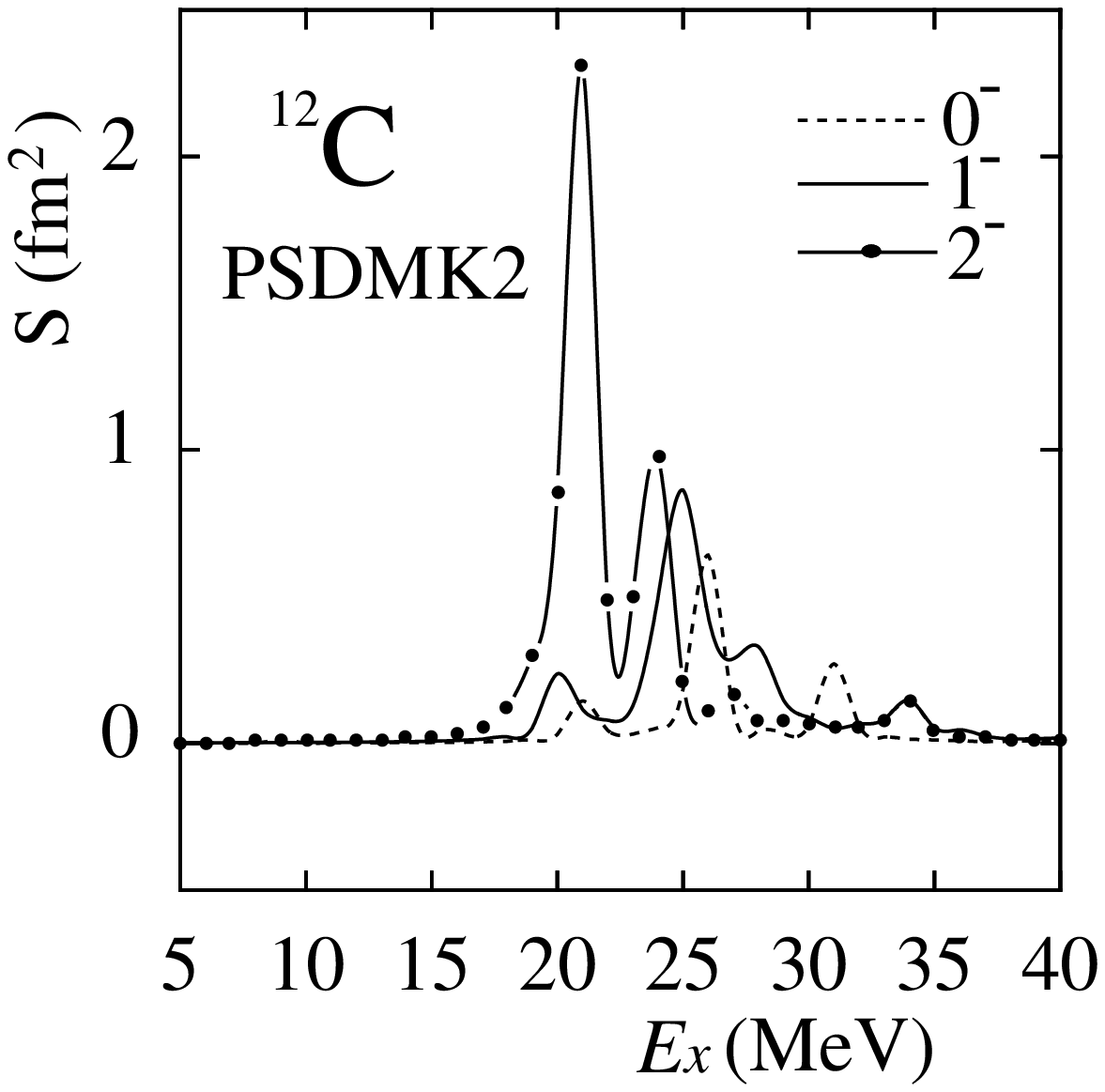}
\vspace*{10mm}
\caption{Spin-dipole strengths in $^{12}$C for the SFO and PSDMK2
Hamiltonians.  The strengths are folded by a Lorenzian with the width 
of 1 MeV. 
\label{fig:fig3}}
\end{figure*}

We focus here on reactions on $^{12}$C, as the Gamow-Teller 
transition to the ground states of $^{12}$N and $^{12}$B are
studied quite well \cite{Wilk,SFO03}. 
Charge-exchange reactions as well as neutral current reactions
induced by DAR neutrinos are investigated.
The electron neutrinos produced from the DAR pions and the $\mu^{+}$ 
decay have average
energy of about 35 MeV with an upper limit at 52.8 MeV. 
The reactions are induced dominantly by the 
axial-vector current. Contributions from the vector current are 
rather small but not negligible.  

The multipole expansions of the reaction cross sections induced 
by $\nu$ or $\bar{\nu}$ are given as follows \cite{DW}, 

\begin{eqnarray}
& &\left(\frac{d\sigma}{d\Omega}\right)_{\frac{\nu}{\bar{\nu}}} =
\frac{G^2\epsilon k}{4\pi^2}\frac{4\pi}{2J_i+1} 
\{\sum_{J=0}^{\infty} \{(1+\vec{\nu}\cdot\vec{\beta})\nonumber\\ 
& &\mid\langle J_f \parallel M_J \parallel J_i\rangle\mid^2 
+ [1-\hat{\nu}\cdot\vec{\beta}+2(\hat{\nu}\cdot\hat{q})(\hat{q}
\cdot\vec{\beta})]\nonumber\\
& &\mid\langle J_f\parallel L_J \parallel J_i\rangle
\mid^2
- \hat{q}\cdot(\hat{\nu}+\vec{\beta}) 2 Re \langle J_f \parallel
L_J \parallel J_i\rangle \nonumber\\
& &\langle J_f \parallel M_J \parallel J_i
\rangle^{\ast}\} 
+ \sum_{J=1}^{\infty} \{[1-(\hat{\nu}\cdot\hat{q})(\hat{q}\cdot
\vec{\beta})]\nonumber\\
& &(\mid\langle J_f \parallel T_{J}^{el} \parallel J_i
\rangle\mid^2 + \mid\langle J_f \parallel T_{J}^{mag} \parallel
J_i\rangle\mid^2 \nonumber\\
& &\pm \hat{q}\cdot(\hat{\nu}-\vec{\beta}) 2 Re [\langle J_f \parallel
T_{J}^{mag} \parallel J_i \rangle \nonumber\\ 
& &\langle J_f \parallel T_{J}^{el}
\parallel J_i \rangle^{\ast}])\} 
\end{eqnarray}
where $\vec{\nu}$ and $\vec{k}$ are neutrino and lepton momenta,
respectively, $\epsilon$ is the lepton energy, $\vec{q}=\vec{k}
-\vec{\nu}$, $\vec{\beta}=\vec{k}/\epsilon$, $\hat{\nu}=\vec{\nu}
/\mid\vec{\nu}\mid$ and $\hat{q}=\vec{q}/\mid\vec{q}\mid$. 
 
For charge-exchange reactions, $G=G_{F}$cos$\theta_C$ with $G_{F}$
the Fermi coupling constant and $\theta_C$ the Cabbibo angle, and
the lepton is electron or positron. 
For neutral current reactions, $G=G_{F}$  and the lepton is 
scattered neutrino.
The cross section is multiplied by the Fermi function 
$F(Z_{f},E_{\ell})$ \cite{Fermi}, where $Z_{f}$ is the charge
of the daughter nucleus and $E_{\ell}$ is the energy of the
charged lepton, in case of charge-exchange reactions.  

In eq. (4), $M_{J}, L_{J}, T_{J}^{el}$ and $T_{J}^{mag}$ are
Coulomb, longitudinal, transverse electric and magnetic multipole
operators for vector and axial-vector currents defined by
\begin{eqnarray}
\langle\vec{p'}\mid J_{\mu}\mid \vec{p}\rangle &=& i\bar{u}(\vec{p'})
[F_{1}^{V}\gamma_{\mu} + F_{2}^{V}\sigma_{\mu\nu}q_{\nu}]\tau_{\mp}
u(\vec{p})\nonumber\\
\langle\vec{p'}\mid J_{\mu}^{5}\mid \vec{p}\rangle &=& i\bar{u}
(\vec{p'}) [F_{A}\gamma_{5}\gamma_{\mu} -iF_{P}\gamma_{5}q_{\mu}]
\tau_{\mp}u(\vec{p})
\end{eqnarray}
for $(\nu, \ell^{-})$ and $(\bar{\nu}, \ell^{+})$ reactions.
Their matrix elements are given in ref.\cite{DW}. 
$F_{1}^{V}$ and $F_{2}^{V}$ are isovector electromagnetic form
factors, $F_{A}$ is the axial-vector form factor with $F_{A}$
($q_{\mu}^{2}$=0) =$g_{A}$, and $F_{P}$ is the induced 
pseudoscalar form factor. Here, we consider vanishing scalar and 
tensor couplings.  
In the extreme relativistic limit, when
the lepton mass can be neglected, the pseudoscalar coupling in the
axial-vector current does not contribute to the neutrino reaction
cross sections\cite{DW}. The induced pseudoscalar terms, 
therefore, can be safely neglected in the present calculations,
where only electrons and positrons are treated except for 
neutrinos as the leptons and neutrino energies are high enough
compared to the electron mass. 

Eq. (4) with the multipole operators obtained for the neutral
current,
\begin{equation}
J_{\mu}^{N} = J_{\mu}^{A_3} + J_{\mu}^{V_3} -2 \mbox{sin}^{2}\theta_{W}
J_{\mu}^{\gamma}
\end{equation}
with 
\begin{eqnarray}
J_{\mu}^{\gamma} &=& J_{\mu}^{S} + J_{\mu}^{V_3} \nonumber\\ 
J_{\mu}^{A_3} &=& i\bar{u}(\vec{p'}) [F_{A}\gamma_{5}\gamma_{\mu}
-iF_{P}\gamma_{5}q_{\mu}] \frac{\tau_{3}}{2} u(\vec{p})\nonumber\\
J_{\mu}^{V_3} &=& i\bar{u}(\vec{p'}) [F_{1}^{V}\gamma_{\mu} +
F_{2}^{V}\sigma_{\mu\nu}q_{\nu}] \frac{\tau_3}{2} u(\vec{p})
\nonumber\\
J_{\mu}^{S} &=& i\bar{u}(\vec{p'}) [F_{1}^{S}\gamma_{\mu} +
F_{2}^{S}\sigma_{\mu\nu}q_{\nu}] \frac{1}{2} u(\vec{p}),  
\end{eqnarray}
applies also to ($\nu, \nu'$) and ($\bar{\nu}, \bar{\nu}'$) 
reactions. Here, $\theta_{W}$ is the Weinberg angle, and 
$F_{1}^{S}$ and $F_{2}^{S}$ are isoscalar
electromagnetic form factors. As mentioned above,
the contributions from the pseudoscalar coupling ($F_{P}$)
vanish.   

First, we show results of cross sections for the exclusive
reaction $^{12}$C ($\nu_{e}, e^{-}$) $^{12}$N ($1_{g.s.}^{+}$)
induced by DAR neutrinos. Calculated cross sections obtained
by using the SFO and the PSDMK2 \cite{OXB,MK,SFO03} shell model 
Hamiltonians within the configuration space including up to 
2$\hbar\omega$ excitations are given in Table I as well as the
observed values \cite{LSND,KARMEN}.   
Harmonic oscillator wave functions with a size parameter $b$
=1.64 fm are used. The axial electric dipole ($E_{5}1$) and 
the magnetic dipole ($M1$) terms contribute to the Gamow-Teller
transition. There are also contributions from the axial Coulomb
and longitudinal dipole ($C_{5}1$ and $L_{5}1$) terms, but 
they are rather small. The bare axial vector coupling constant,
$g_{A}$ = -1.263 and an effective one with $g_{A}^{eff}$ = 0.95
$g_{A}$ are used. The latter one reproduces the experimental
$B(GT)$ value for the transition to $^{12}$N ($1_{g.s.}^{+}$).
While the SFO Hamiltonian gives larger values of the cross
section than those obtained by other conventional shell model
Hamiltonians \cite{VACSG,HT}, the calculated cross sections 
are found to be consistent with the experimental ones within
the experimental errors \cite{LSND,KARMEN}. 
A no core shell model (NCSM) calculation gives a smaller 
value of 6.80$\times$10$^{-42}$ cm$^{2}$ for the cross section 
\cite{Navr}, while a
continuum random phase approximation (CRPA) method gives
a larger value of 13.88$\times$10$^{-42}$ cm$^{2}$ \cite{Lang}. 



Next, we show in Table II calculated results of the cross sections
for exclusive neutral current reactions on $^{12}$C, that is,
($\nu_{e}$, $\nu_{e}'$), ($\bar{\nu}_{\mu}$, $\bar{\nu}_{\mu}'$)
and ($\nu_{\mu}$, $\nu_{\mu}'$) reactions leading to the 1$^{+}$
($T$=1, 15.1 MeV) state of $^{12}$C induced by the DAR neutrinos.
Experimental value of the cross section for the sum of the 
($\nu_{e}$, $\nu_{e}'$) and ($\bar{\nu}_{\mu}$, $\bar{\nu}_{\mu}'$)
reactions is available \cite{KARMEN} (see Table II). 
The calculated cross section obtained for the SFO Hamiltonian is
found to be close to the observed value, while that for the PSDMK2
Hamiltonian is smaller than the experimental one about by 20$\%$.
Note that the $B(GT)$ value obtained for the PSDMK2 Hamiltonian with
the 0$-$2 $\hbar\omega$ configuration space is smaller than the 
observed one by 16$\%$ \cite{SFO03}.  
Experimental value of the ($\nu_{\mu}, \nu_{\mu}'$) reaction cross
section \cite{KARM2} is also found to be consisitent with the calculated 
value for the SFO Hamiltonian whereas it is a bit larger than the
calculated value for the PSDMK2 Hamiltonian.   

Finally, we show in Table III calulated results of the cross 
section for $^{12}$C ($\nu_{e}, e^{-}$) $^{12}$N$^{\ast}$ 
reaction leading to excited states of $^{12}$N obtained by the
shell models with the configuration space including up to 
3$\hbar\omega$ excitations. 
The multipolarities up to $J$=3 are included here. 
The contributions from the spin-dipole transitions to 
the $0^{-}$, $1^{-}$ and $2^{-}$ states are dominant. 
There are also some contributions from other multipolarities,
$2^{+}$, $3^{-}$, $3^{+}$ and $0^{+}$ as well as $1^{+}$ 
except for the ground state. 
The SFO Hamiltonian gives closer energy levels for the
negative parity states than the PSDMK2 Hamiltonian. 
The excitation energies of the first $0^{-}$, $1^{-}$, $2^{-}$
and $3^{-}$ states with $T$=1 are 20.304 (21.086) MeV, 19.053 (20.137)
MeV, 17.823 (18.881) MeV and 19.087 (20.128) MeV, respectively, 
for the SFO (PSDMK2) case, while experimental values are 17.230 MeV 
($1^{-}$), 16.570 MeV ($2^{-}$) and 18.350 MeV ($3^{-}$). 
The spin dipole strengths obtained by the SFO and PSDMK2 Hamiltonians 
are shown in Fig. 3.  Calculated strengths  
summed up to the excitation energy of $E_{x}$ = 50 MeV 
are 1.79 (1.78) fm$^{2}$, 4.33 (4.12) fm$^{2}$ and 7.19 (7.03) fm$^{2}$ 
for the $0^{-}$, $1^{-}$ and $2^{-}$ states, respectively, for the
SFO (PSDMK2) Hamiltonian. 
The centroid energies defined by the energy-weighted sum divided by
the non-energy-weighted sum of the strength are calculated to be
25.9 (26.8) MeV, 25.3 (26.2) MeV and 21.5 (22.7) MeV for the
SFO (PSDMK2) case for the $0^{-}$, $1^{-}$ and $2^{-}$ states, 
respectively. The strength by the SFO Hamiltonian is shifted toward
lower energy region by about 1 MeV compared to the PSDMK2 Hamiltonian
while the total strength is increased only about by 3$\%$.         

Shell model calculations give larger cross sections than
the observed values \cite{LSND,KARM3} except for one by Hayes 
and Towner \cite{HT} where Woods-Saxon wave functions are
used instead of harmonic oscillator wave functions. 
A CRPA calculation \cite{Lang} gives a cross section
close to the experiment. 
Effective axial-vector coupling constants with quenching
factors, $g_{A}^{eff}/g_{A}$ =0.7 and 0.75, are adopted 
for the SFO and PSDMK2 Hamiltonians, respectively. 
Shell model calculations with these quenching factors reproduce
the experimental cross section. 

There is an observational indication from electron sacttering and
($p, n$) reaction data that the spin-dipole strength in the 2$^{-}$ 
($T$=1) state in $^{12}$C ($^{12}$N) at $E_{x}$ =19.40 (4.14) MeV 
is considerably quenched by a factor of about 2 \cite{Drake,Yang,
Gaard}. Note that the 2$^{-}$ state exhausts about 60$\%$ of the total
spin-dipole strength for 2$^{-}$ states (see also ref. \cite{Gaard}).
This results in more importance of the 2$^{-}$ state in the cross 
section, that is, about 75$\%$ of the cross section for 
$^{12}$C ($\nu_{e}, e^{-}$) $^{12}$N$^{\ast}$ (2$^{-}$) induced by
DAR neutrinos comes from the 2$^{-}$ state at 4.14 MeV due to the 
neutrino energy cut-off at 52.8 MeV. 
The $M2$ form factor for $^{12}$C ($e, e'$) $^{12}$C (2$^{-}$, $T$=1, 
19.40 MeV) was obtained in the momentum transfer region of $q$ 
=0.3$\sim$1.0 fm$^{-1}$ \cite{Drake}. The observed form factor is
found to be consistent with a large quenching of the spin $g$ factor:
$g_{s}^{eff}/g_{s}$ = 0.70$\pm$0.05 (0.75$\pm$0.05) for the SFO (PSDMK2)
hamiltonian. This was also pointed out in ref. \cite{Gaard}, where 
$g_{s}^{eff}/g_{s}$ = 0.65 was obtained for the Cohen-Kurath Hamiltonian.
The ($p, n$) and ($d$, $^{2}$He) reaction data support similar order of 
large quenching factors \cite{Gaard,Yang,Okam}. More experimental 
investigation is important and necessary to get systematic information
on the nature of quenching of the spin-dipole strength.
Quenching due to the coupling to many-particle many-hole states at 
high excitation energies could be larger for the spin-dipole states
which lie above the Gamow-Teller state because of smaller 
energy-difference denominator.

In case of 2$^{+}$ ($T$=1) state, ($p, n$) and ($p, p'$) reaction data
indicate that the transition strength to the 2$^{+}$ ($T$=1) state
in $^{12}$N ($^{12}$C) at $E_x$ = 0.96 (16.11) MeV is quenched by a
factor of about 2 \cite{Rapapo, Comf}. It was also found in 
ref. \cite{SzSag} that in the electric dipole transitions in $^{12}$C
the reduction of the calculated cross section by a multiplying factor
of 0.7 was necessary in order to obtain quantitative agreement with 
the available experimental cross section \cite{Pywel}.  This suggests
the importancce of the coupling to many-particle many-hole states with
excitations larger than 3$\hbar\omega$.  We thus get several supports
from observations for the necessity for the large quenching of $g_{A}$ 
in the inclusive reactions.

\subsection{REACTIONS ON $^{12}$C INDUCED BY SUPERNOVA NEUTRINOS}

We study charge-exchange and neutral current reactions on 
$^{12}$C induced by the supernova neutrinos. Fermi distribution 
functions are employed for the spectra of the supernova neutrinos. 
The value of the chemical potential is set to be zero. Average
energies of supernova neutrinos are about 10 MeV, 15 MeV and 
15$\sim$25 MeV for $\nu_{e}$, $\bar{\nu}_{e}$, $\nu_{\mu}$ and 
$\nu_{\tau}$, respectively \cite{ASTR}.  
The neutrino temperature of the Fermi distribution is about
one-third of the average energy. 

\begin{figure*}[tbh]
\vspace{-10mm}
\hspace{-10mm}
\includegraphics[scale=0.48]{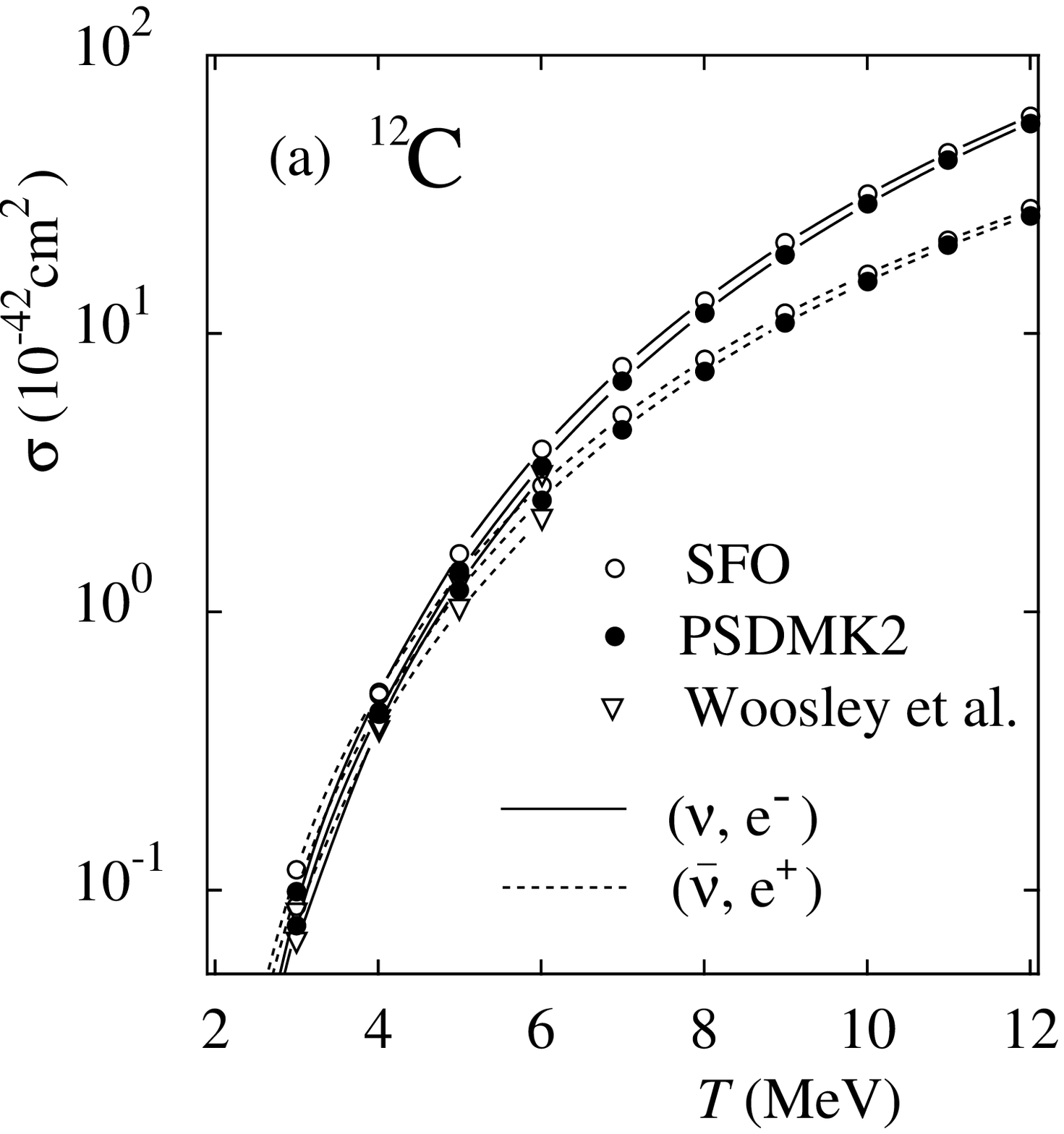}
\vspace{-10mm}
\hspace{-0mm}
\includegraphics[scale=0.48]{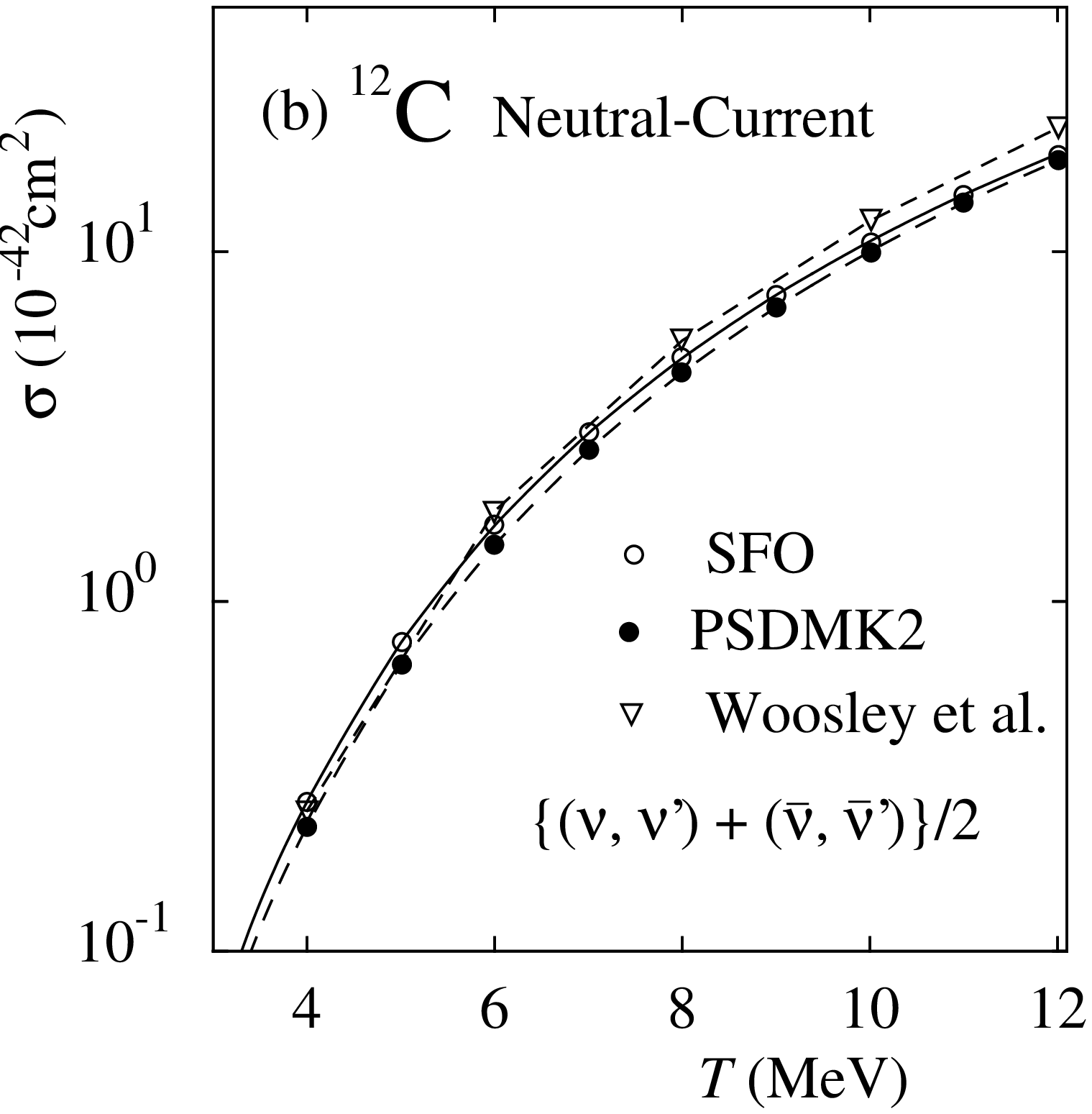} 
\caption{Calculated cross sections for neutrino $^{12}$C reactions
induced by supernova neutrinos with temperature $T$ obtained by using 
the SFO and PSDMK2 Hamiltonians. 
Both (a) the charge-exchange ($\nu_{e}, e^{-}$) and ($\bar{\nu}_{e},
e^{+}$) reactions, and (b) the neutral current reactions are treated. 
Average values of the ($\nu, \nu'$) and ($\bar{\nu}, \bar{\nu}'$)
cross sections are shown for the neutral current reactions.
Previous calculations of ref. \cite{WHHH} are also given.   
\label{fig:fig4}}
\end{figure*}

\begin{figure*}[tbh]
\vspace{-10mm}
\hspace{-10mm}
\includegraphics[scale=0.48]{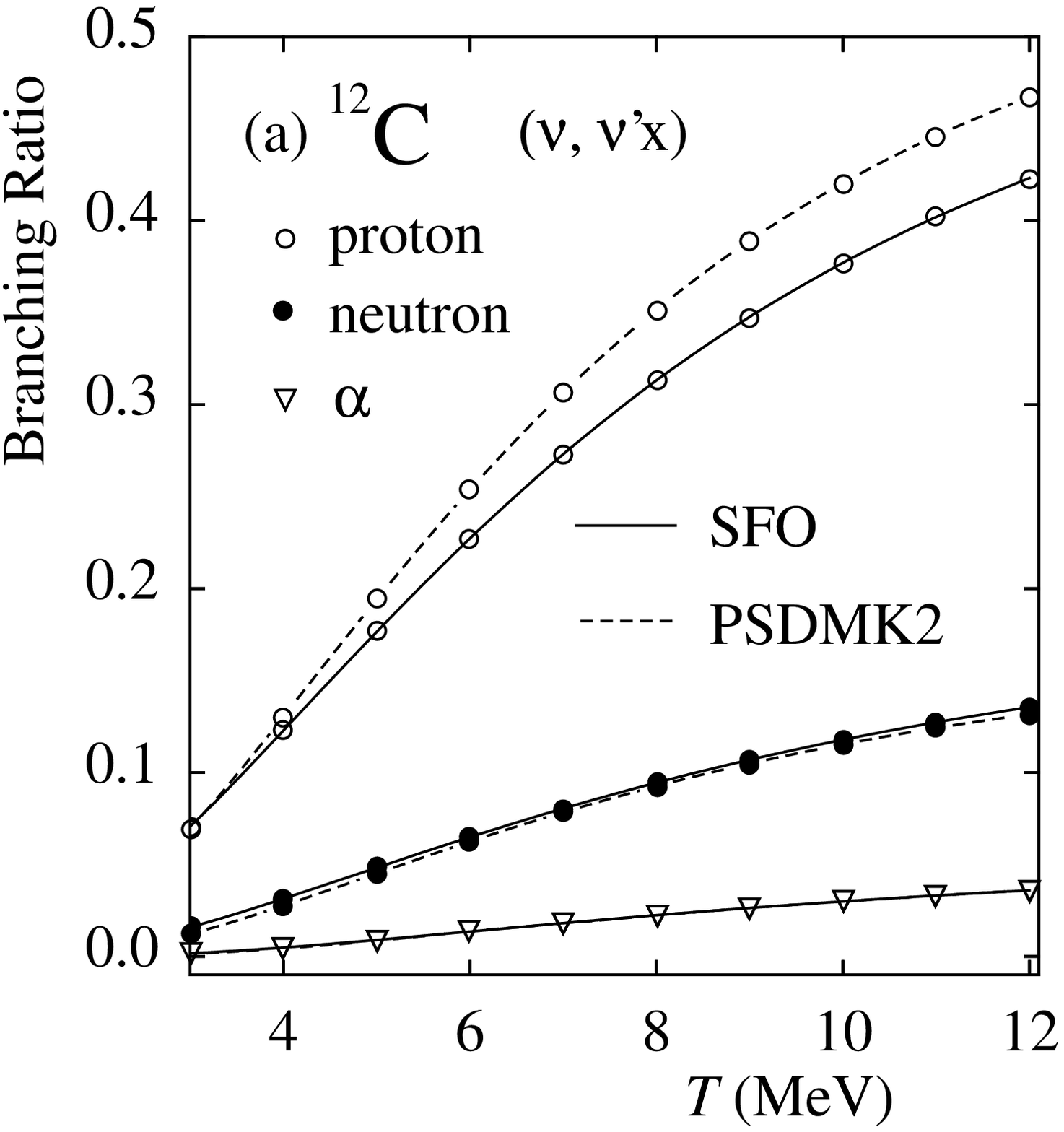}
\vspace{-10mm}
\hspace{3mm}
\includegraphics[scale=0.48]{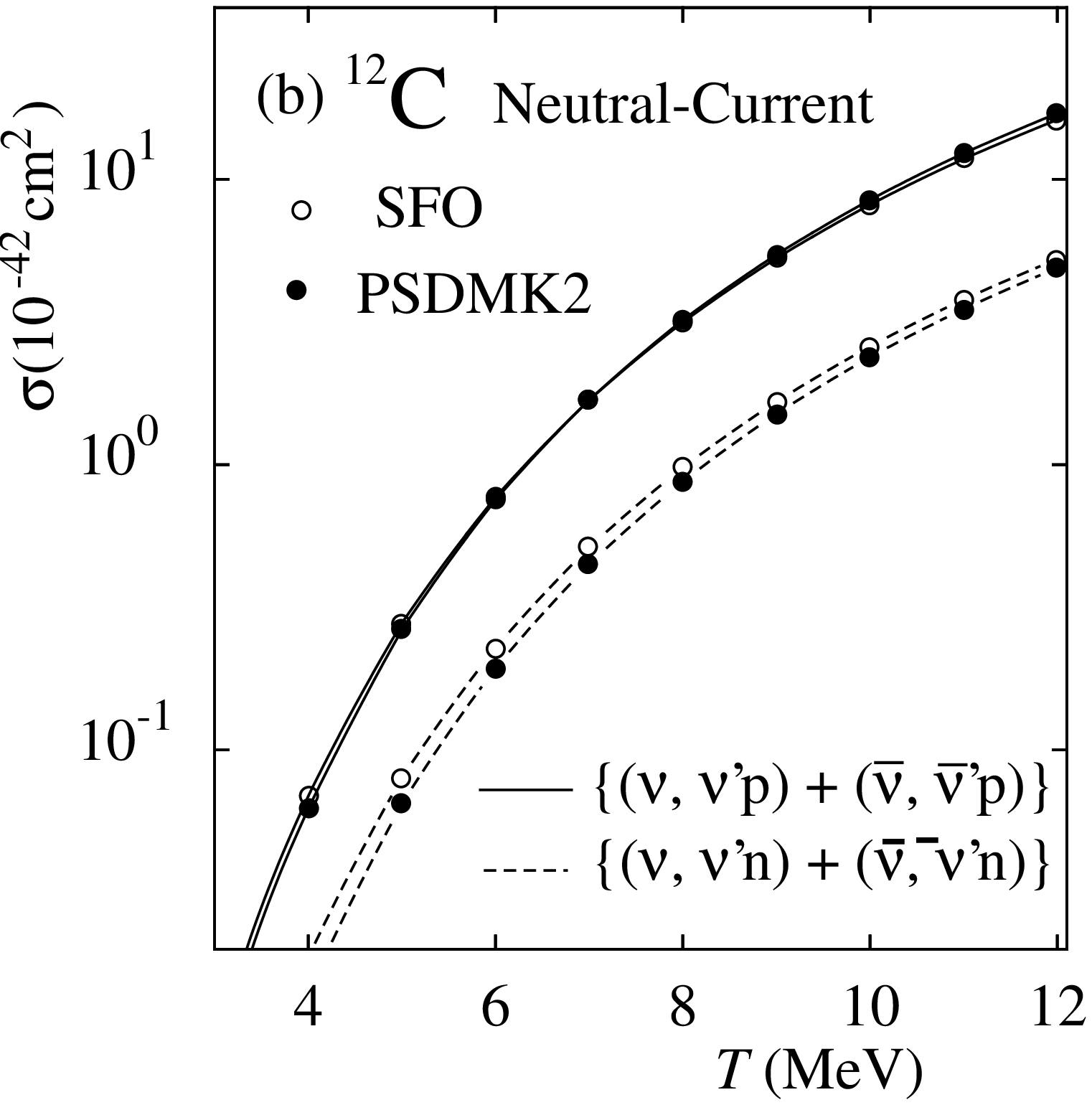}
\caption{(a) Branching ratios for proton, neutron and $\alpha$ 
emission channels for neutral current reactions on $^{12}$C obtained by
the Hauser-Feshbach theory. 
(b) Calculated cross sections for proton and neutron knock-out channels
obtained by using the SFO and PSDMK2 Hamiltonians.
\label{fig:fig5}}
\end{figure*}

\begin{figure*}[tbh]
\vspace{-35mm}
\hspace{-10mm}
\includegraphics[scale=0.53]{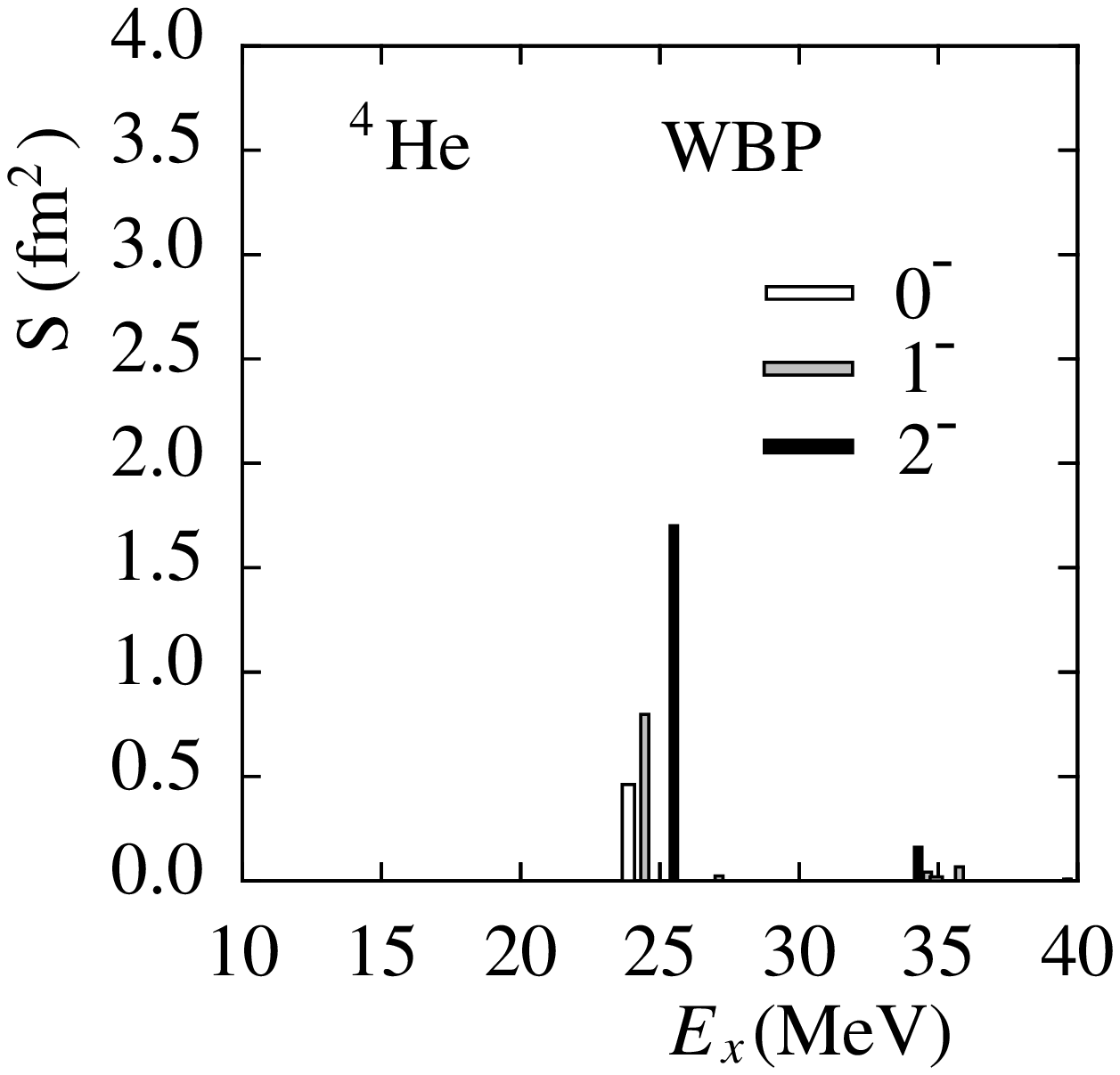}
\vspace{-35mm}
\hspace{-15mm}
\includegraphics[scale=0.53]{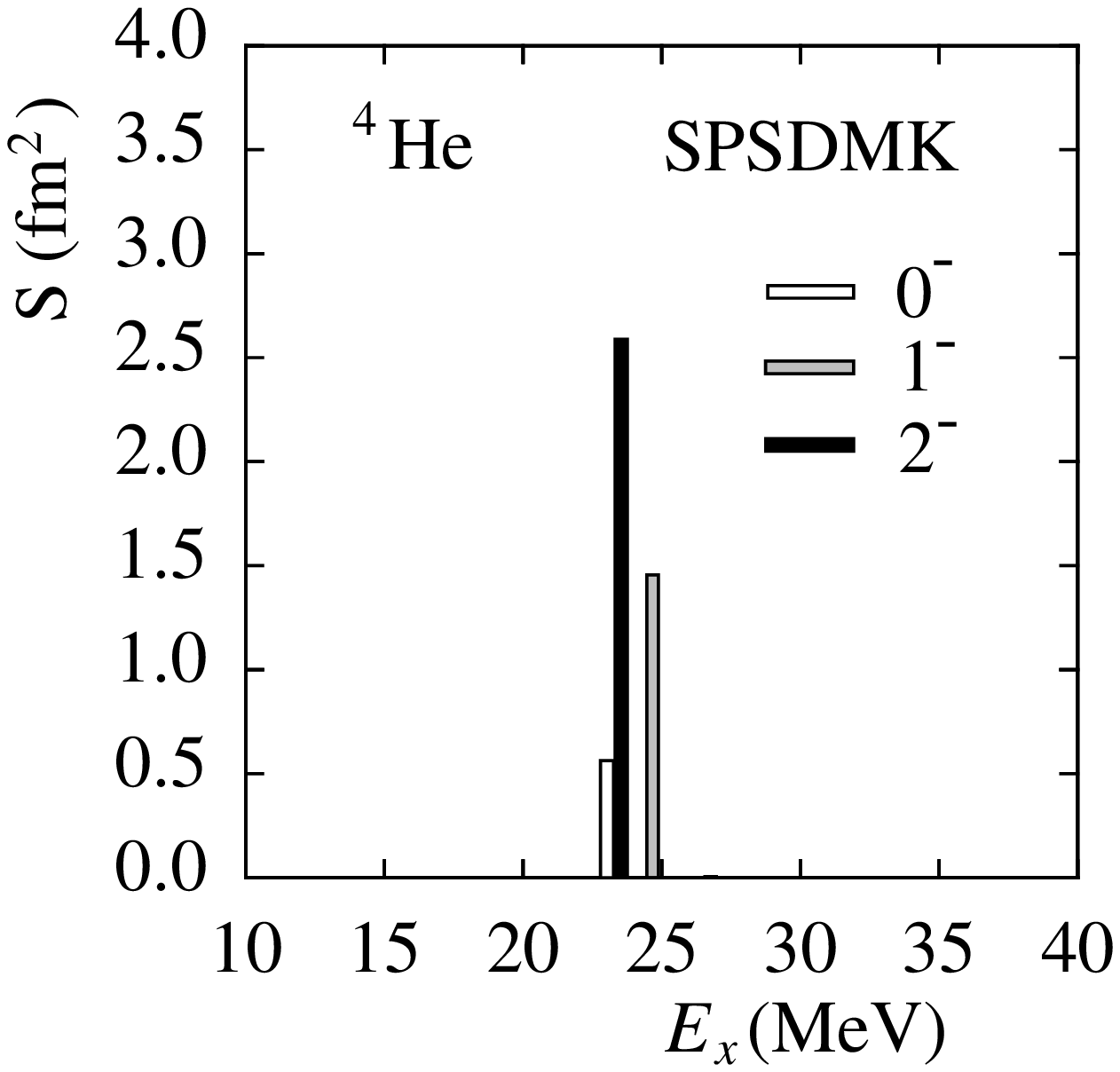}
\vspace*{10mm}
\caption{Spin-dipole strengths in $^{4}$He for the WBP and SPSDMK
Hamiltonians
\label{fig:fig6}}
\end{figure*}

Calculated cross sections for supernova neutrinos with
temperature $T$ = 2$\sim$12 MeV are shown in Fig. 4 for the
SFO and PSDMK2 Hamiltonians. The axial-vector coupling
constants which reproduce the experimental ($\nu_{e}, e^{-}$) 
cross section for the DAR neutrinos are adopted. The values of
$g_{A}^{eff}/g_{A}$ are 0.95 and 0.70 for the exclusive reaction
and the transitions to the excited states, respectively, in case
of the SFO Hamiltonian. Those employed for the PSDMK2 Hamiltonian
are 1.0 for the exclusive reaction and 0.75 for the transitions
to the excited states, respectively. 
We will use these values for $g_{A}^{eff}$ hereafter. 
As for the neutral current reactions, contributions from the
isoscalr transitions are not included in the calculations as
they are quite small.  
Calculated cross sections for the SFO are enhanced 
compared with those for the PSDMK2 in both charge-exchange
and neutral current reactions. The charge-exchange
reaction cross sections are also enhanced compared with the
previous calculations by Woosley et al.\cite{WHHH}, in which
the configurations are restricted to up to 1$\hbar\omega$ 
excitations with $g_{A}^{eff}/g_{A}$ =0.7 for the excitations
of the negative-parity states.

Branching ratios from each excited level are calculated for
decay channels involving neutron, proton, $\alpha$ and $\gamma$
by the Hauser-Feshbach statistical model \cite{HF}. All the 
levels obtained by the present shell model calculations are
adopted as levels in the decaying and daughter nuclei with
specific isospin assignments.  

The particle transmission coefficients are calculated by the
optical model with conventional potentials \cite{WG,AVR94}
at selected grid energies, and they are interpolated by using
a spline interpolation. Weights proportional to the square of
the isospin CG-coefficients are multiplied to the transmission 
coefficients obtained by the optical model to account for the
isospin conservation. We ignored any isospin mixing which
may be significant for some of the light nuclei. 
The $\gamma$-transmission coefficients are calculated with
a simple Brink's formula. The E1 and M1 parameters were taken
from RIPL-2 database \cite{RIPL2}. The $\gamma$-cascade in
the initial excited nuclei and subsequent decay were fully
considered.  

\begin{figure*}[tbh]
\vspace{-35mm}
\hspace{-15mm}
\includegraphics[scale=0.53]{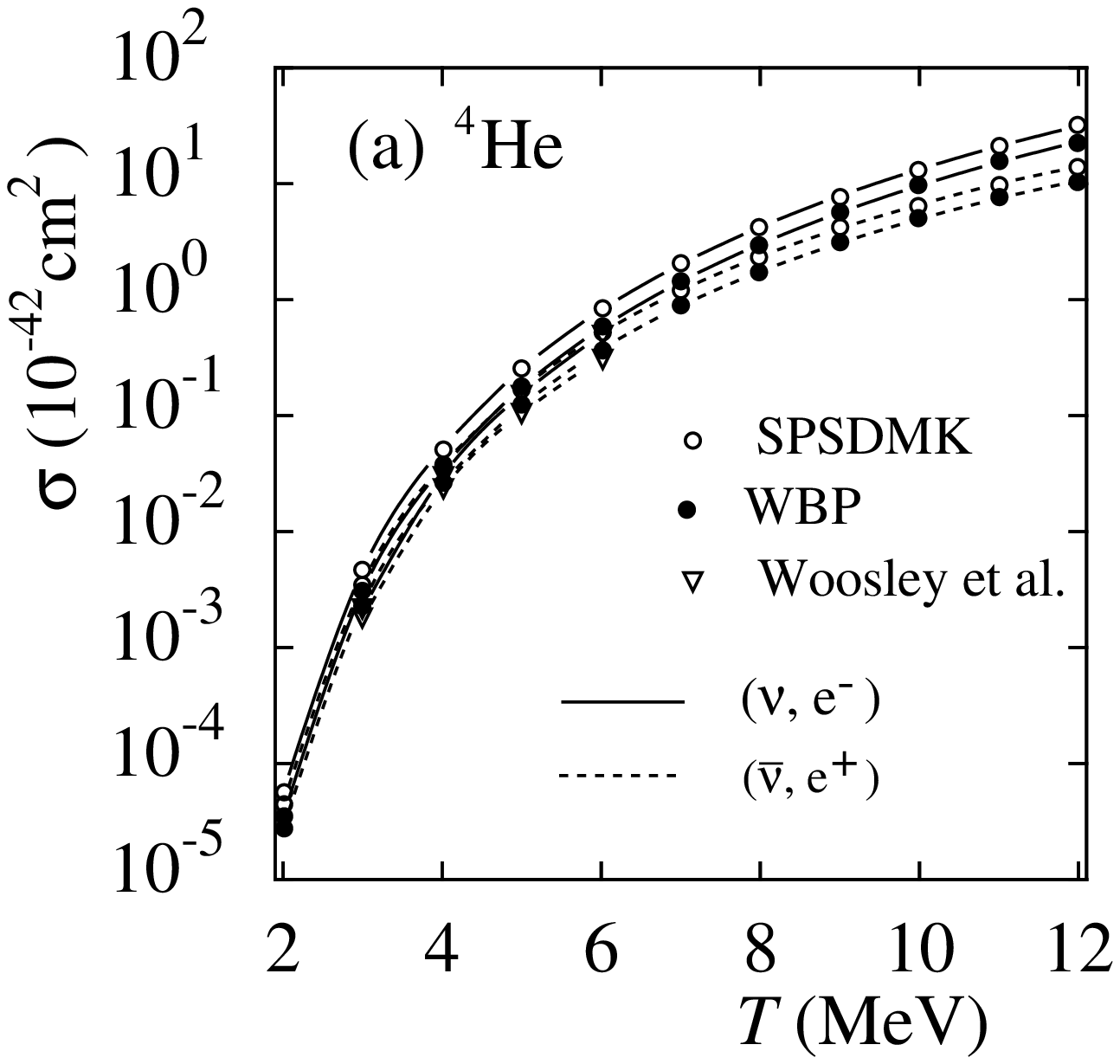}
\vspace{-35mm}
\hspace{-15mm}
\includegraphics[scale=0.53]{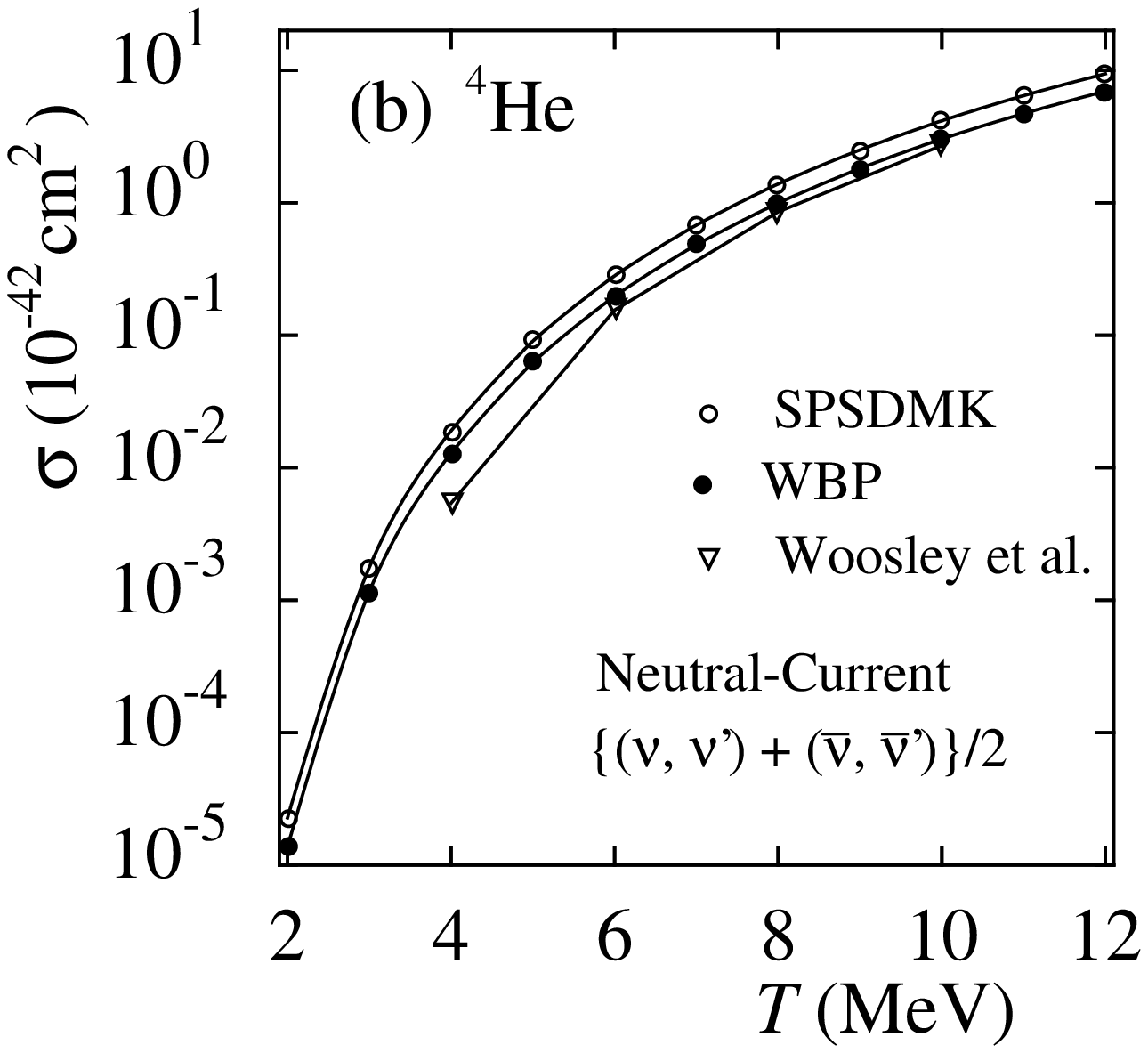}
\vspace{2.8cm}
\caption{Calculated (a) charge-exchange and (b) neutral current 
reaction cross sections for $\nu$-$^{4}$He reactions obtained by
using the WBP and SPSDMK Hamiltonians. Previous calculations of
ref. \cite{WHHH} are also shown.
\label{fig:fig7}}
\end{figure*}

Calculated branching 
ratios as well as the proton and neutron emission cross
sections are shown in Fig. 5 for the neutral current reactions. 
The branching ratios for the proton and neutron
emissions obtained by the PSDMK2 Hamiltonian
are close to those of ref. \cite{WHHH}. 
The branching ratios for the proton emissions depend on
the Hamiltonians, SFO or PSDMK2, while
the neutron emission cross sections are found to be enhanced
for the SFO case.     

Neutral current reactions, $^{12}$C ($\nu, \nu'$p) $^{11}$B and
$^{12}$C ($\nu, \nu'$n) $^{11}$C ($\beta^{+}$) $^{11}$B, are 
important for the production of $^{11}$B in the He-C layer and O-rich 
layer during supernova explosions. The effects of the reactions
on the abundance of $^{11}$B in the supernovae will be 
discussed in sect. IV.

\subsection{REACTIONS ON $^{4}$He}

We treat here the $^{4}$He nucleus for the study of neutrino-nucleus 
reactions. The reaction cross sections on the nucleus has an 
important role in determining abundances of light elements such as 
$^{7}$Li and $^{11}$B during supernova explosions.  

The $\nu$-$^{4}$He reactions are induced dominantly by excitations
of spin-dipole states. The Warburton-Brown (WBP) \cite{WB} and 
Millener-Kurath (SPSDMK) \cite{MK,OXB} Hamiltonians are used for
the shell model calculations of $^{4}$He with configurations
including up to 3$\hbar\omega$ excitations. In the SPSDMK 
interaction, the Cohen-Kurath interaction, (8-16)POT \cite{CK}, 
is used for the $p$-shell part while the Millener-Kurath interaction
is used for the cross-shell matrix elements between 0$s$ and 0$p$
as well as 0$p$ and $1s0d$ orbits. The $sd$-shell part is the
Preedom-Wildenthal interaction \cite{PW}, and all others are
Kuo's renormalized G-matrices \cite{Kuo}. 

Calculated spin-dipole strengths are shown in Fig. 6. Harmonic
oscillator wave functions with a size parameter $b$ = 1.38 fm 
are used. The strength is more fragmented in the case of the WBP
Hamiltonian. The summed strengths are 3.34 fm$^{2}$ for the
WBP and 4.71 fm$^{2}$ for the SPSDMK Hamiltonians, respectively,
up to the excitation energy of $E_{x}$ = 50 MeV.

\begin{figure*}[tbh]
\includegraphics[scale=0.55]{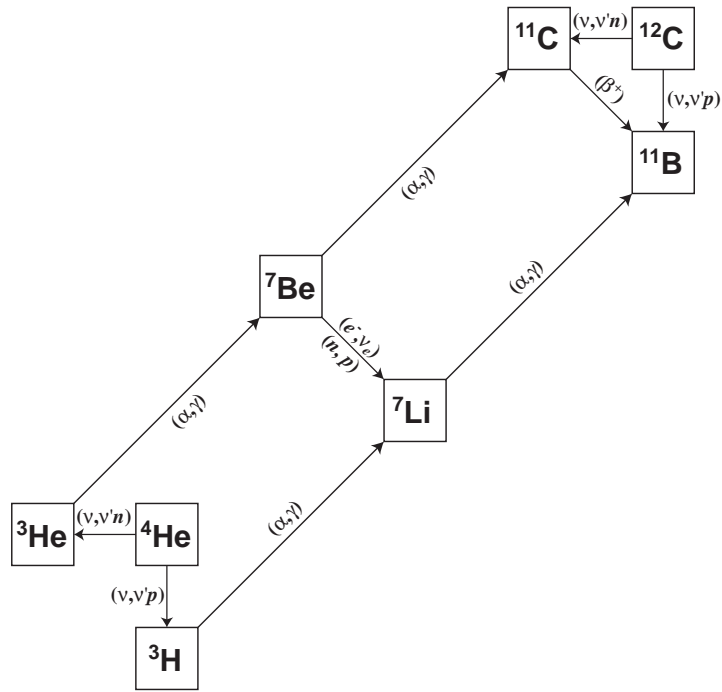}
\caption{Nucleosynthesis path of light elements $^{7}$Li and $^{11}$B
during supernova explosions. 
\label{fig:fig8}}
\end{figure*}

Calculated cross sections for the charge-exchange and neutral
current reactions are shown in Fig. 7 for the supernova 
neutrinos with $T$ =2$\sim$12 MeV. The bare $g_{A}$ is employed.
The cross sections for the SPSDMK case are found to be larger 
than those for the WBP case. They are both enhanced compared
to the previous calculations \cite{WHHH}, where Sussex matrix
elements \cite{Elliot} are used for the effective interaction 
with a larger harmonic oscillator size parameter of $b$= 1.5
fm. 

A recent microscopic ${\it ab}$ ${\it initio}$ calculation of the
neutral current reaction on $^{4}$He with a realistic
nucleon-nucleon interaction, $AV8'$ \cite{AV8P}, predicts
cross sections with a steeper dependence on the neutrino
temperature than the shell model calculations \cite{Gazit}. 
At $T$= 10 MeV, the calculated cross section in ref. \cite{Gazit}
is close to that obtained by the WBP Hamiltonian, while at
$T$= 12 MeV it is enhanced by about 16$\%$. At $T$= 8 MeV,
on the other hand, the WBP Hamiltonian predicts a larger 
cross section by about 15$\%$. 

The neutral current reactions, $^{4}$He ($\nu, \nu'$p) $^{3}$H
and $^{4}$He ($\nu, \nu'$p) $^{3}$He, are important for the
production of $^{7}$Li through $^{3}$H ($\alpha, \gamma$) $^{7}$Li
and $^{3}$He ($\alpha, \gamma$) $^{7}$Be ($e^{-}, \nu_{e}$) $^{7}$Li
processes in the He-C layer during supernova explosions. 
The reactions are also important for the production $^{11}$B as
the abundance of $^{7}$Li affects the production of $^{11}$B
through the process $^{7}$Li ($\alpha, \gamma$) $^{11}$B.  
  
\begin{table*}
\caption{\label{tab:table4}
Production yields of $^{7}$Li and $^{11}$B 
in a supernova explosion model}
\begin{tabular}{ c|c|c } \hline
Hamiltonians & $M$($^{7}$Li)/$M_{\bigodot}$ & $M$($^{11}$B)
/$M_{\bigodot}$ \\\hline
WBP + SFO & 3.06 $\times$ 10$^{-7}$ & 7.51 $\times$ 10$^{-7}$ \\\hline
SPSDMK + PSDMK2 & 4.24 $\times$ 10$^{-7}$ & 9.38 $\times$ 10$^{-7}$
\\\hline
HW92 & 2.36 $\times$ 10$^{-7}$ & 6.29 $\times$ 10$^{-7}$ \\\hline
\end{tabular}
\end{table*}

\begin{table*}
\caption{\label{tab:table5}
Dependence of the production yields of $^{7}$Li and $^{11}$B
on neutrino temperatures and total neutrino energy.
Reaction cross sections by (a) WBP and SFO Hamiltonians as well 
as (b) SPSDMK and PSDMK2 Hamiltonians for $^{4}$He
and $^{12}$C, respectively, are used.}
\begin{tabular}{ c|c|c|c|c|c|c|c } \hline
Hamiltonians & Neutrino model & $T_{\nu_{e}}$ & $T_{\bar{\nu}_{e}}$ & 
$T_{\nu_{\mu,\tau}}$ & $E_{\nu}$  & $M$($^{7}$Li) & 
$M$($^{11}$B) \\
 & & (MeV) & (MeV) & (MeV) & ($\times10^{53}$ erg) & 
($M_{\bigodot}$) & ($M_{\bigodot}$) \\\hline
(a) WBP + SFO & Low $T_{\nu}$, High $E_{\nu}$ & 3.2 & 4.1 & 5.0 & 3.53 
 & 1.51$\times10^{-7}$ & 3.59$\times10^{-7}$ \\
 & Med $T_{\nu}$, Med $E_{\nu}$ & 3.2 & 4.8 & 5.8 & 3.0 & 
2.62$\times10^{-7}$ & 6.36$\times10^{-7}$ \\
 & High $T_{\nu}$, Low $E_{\nu}$  & 3.2 & 5.0 & 6.4 & 2.35 &
3.13$\times10^{-7}$ & 7.45$\times10^{-7}$ \\\hline
(b) SPSDMK + PSDMK2 & 
Low $T_{\nu}$, High $E_{\nu}$ & 3.2 & 4.0 & 4.8 & 3.53 &
1.76$\times10^{-7}$ & 3.55$\times10^{-7}$ \\
 & Med $T_{\nu}$, Med $E_{\nu}$ & 3.2 & 4.6 & 5.6 & 3.0 &
3.07$\times10^{-7}$ & 6.57$\times10^{-7}$ \\
 & High $T_{\nu}$, Low $E_{\nu}$  & 3.2 & 5.0 & 6.0 & 2.35 &
3.41$\times10^{-7}$ & 7.35$\times10^{-7}$ \\\hline
\end{tabular}
\end{table*}

\section{ABUNDANCES OF $^{7}$Li AND $^{11}$B 
DURING SUPERNOVA EXPLOSIONS}

The enhancement of $\nu$-$^{4}$He and $\nu$-$^{12}$C reaction cross
sections affects the abundances of $^{7}$Li and $^{11}$B in 
the nucleosynthesis process during supernova explosions. 
The nucleosynthsis path of light elements is shown in Fig. 8.
The neutral current reactions, $^{12}$C ($\nu, \nu'$p) $^{11}$B
and $^{12}$C ($\nu, \nu'$n) $^{11}$C, are important for the
production of $^{11}$B. If these cross sections are enhanced, 
the abundance of $^{11}$B is increased. 
The reactions, $^{4}$He ($\nu, \nu'$p) $^{3}$H and $^{4}$He
($\nu, \nu'$n) $^{3}$He are important for the production of
$^{7}$Li through $^{3}$H ($\alpha, \gamma$) $^{7}$Li and 
$^{3}$He ($\alpha, \gamma$) $^{7}$Be ($e^{-}, \nu_{e}$) 
$^{7}$Li processes. 
If the $\nu$-$^{4}$He reaction cross sections are enhanced, 
the abundances of both $^{7}$Li and $^{11}$B are increased 
as the abundance of $^{11}$B is affected by that of $^{7}$Li
through $^{7}$Li ($\alpha, \gamma$) $^{11}$B etc. 

In order to investigate the effects of our new reaction cross sections for
$\nu-^4$He and $\nu-^{12}$C on the yields of $^7$Li and $^{11}$B in a
core-collapse supernova, we carry out detailed nucleosynthesis calculation
during supernova explosions.
The supernova explosion model is the same as in \cite{Yoshida,YKH}.
The progenitor structure is adopted from a 16.2 $M_\odot$ presupernova
model corresponding to SN 1987A \cite{Shige}.
The nuclear reaction network consists of 291 species of nuclei.
The luminosity and energy spectra of neutrinos are important for
neutrino-nucleus interactions.
We assume that the neutrino luminosity decays exponentially in time with a
time scale of 3 s and is equally partitioned among three flavors of
neutrinos and antineutrinos.
The neutrino energy spectra are assumed to obey Fermi distributions
with zero-chemical potentials.
We set the total neutrino energy to be $3 \times 10^{53}$ erg and the
neutrino temperatures of $\nu_e$,
$\bar{\nu}_e$, and $\nu_{\mu,\tau}$ and $\bar{\nu}_{\mu,\tau}$ to be
$T_{\nu_e} = 3.2$ MeV, $T_{\bar{\nu}_e} = 5.0$ MeV, and
$T_{\nu_{\mu,\tau}} = 6.0$ MeV, respectively \cite{Yoshida,YKH}, as the 
standard case.

We show in Table IV production yields of $^{7}$Li and $^{11}$B
in the nucleosynsthesis during the supernova explosion obtained
by using the cross sections of the two sets of the shell model
Hamiltonians; one by WBP for $^{4}$He and SFO for $^{12}$C and 
the other by SPSDMK for $^{4}$He and PSDMK2 for $^{12}$C. 
The production yields obtained by using the cross sections
of Hoffman and Woosley (HW92) \cite{HW92} are also given.
Compared to the case by HW92, the abundances of $^{7}$Li and
$^{11}$B are enhanced by a factor of 1.30 and 1.19, respectively,
for WBP+SFO, while they are enhanced more by a factor of 1.79
and 1.49, respectively, for SPSDMK+PSDMK2. Enhancement factors
for $^{7}$Li are larger than those for $^{11}$B, which comes 
from the fact that the cross sections for $^{4}$He are more
enhanced than those for $^{12}$C. 
We find that about 40$\%$ of the production of $^{11}$B is
caused by the $\nu$-$^{12}$C reactions while the other 60$\%$
is due to the $\nu$-$^{4}$He reactions. 


We investigate the production yields of $^7$Li and $^{11}$B by changing
the temperature of $\nu_e$, $\bar{\nu}_e$, and $\nu_{\mu,\tau}$.
We set the following restrictions to the temperature;
(1) $T_{\nu_e} < T_{\bar{\nu}_e} < T_{\nu_{\mu,\tau}}$,
(2) $T_{\bar{\nu}_e} \le 5$ MeV, and
(3) $T_{\nu_{\mu,\tau}}/T_{\bar{\nu}_e} \simeq 1.2$.
We further choose temperature so that the production yield of $^{11}$B
ranges between $3.3 \times 10^{-7} M_\odot$ and
$7.4 \times 10^{-7} M_\odot$ in order to satisfy the supernova contribution
of B abundance in the Galactic chemical evolution \cite{YKH,YTKS}.
The total neutrino energy is varied within the estimated error bar for
the released gravitational binding energy of a proto-neutron star \cite{Latt}.
Calculated results are given in Table V.
Although the production yield of $^{11}$B depends little on the Hamiltonians, 
the production yield of $^7$Li is slightly larger for the case of SPSDMK+PSDMK2
compared with the case of WBP+SFO.

When we take into account the effect of neutrino oscillations, 
charge-exchange
reactions make an additional role to increase both $^7$Li and $^{11}$B
yields.
The production yields prove to be sensitive to the mixing angles,
in particular to $\theta_{13}$, and mass hierarchy \cite{Yoshida}.
This subject with the use of our new reaction cross sections will be
discussed in ref. \cite{YKSC}.

\section{SUMMARY}

Neutrino-nucleus reactions on $^{12}$C induced by DAR neutrinos
and supernova neutrinos are investigated by using a new shell
model Hamiltonian for $p$-shell nuclei, SFO, which takes into
accout important roles of spin-isospin interactions. 

First, the monopole terms of the tensor components of the SFO
interaction are shown to have proper signs, that is, the p-n
interaction is attractive between $j_{\rangle}$ and $j_{\langle}$ 
orbits but repulsive between $j_{\rangle}$ and $j_{\rangle}$ or
$j_{\langle}$ and $j_{\langle}$ orbits. This zigzag structure
of the tensor interaction is pointed out to be important to
realize the proper evolution of effective single particle 
energies toward the drip-lines. For $N$=8 isotones, the shell
gap between the $0p_{1/2}$ and $0p_{3/2}$ orbits is shown to
increase near the neutron-rich side. 

Cross sections of charge-exchange exclusive and inclusive 
reactions on $^{12}$C are, then, obtained for the DAR 
neutrinos with the use of the SFO Hamiltonian, and compared
with experimental values. The exclusive reaction is found
to be well reproduced with $g_{A}^{eff}$ =0.95 $g_{A}$. 
A quenching of $g_{A}$ ($g_{A}^{eff}$ =0.7 $g_{A}$) is 
found to be necessary to explain the cross section for
excited states. 

Charge-exchange and neutral current reactions are studied
also for supernova neutrinos. Branching ratios to proton,
neutron, $\alpha$ and $\gamma$ emission channels are 
calculated by the Hauser-Feshbach theory, and cross sections
for ($\nu, \nu'$p) and ($\nu, \nu'$n) reactions are obtained.
Calculated cross sections are found to be enhanced compared
to those by the PSDMK2 Hamiltonian. 

Neutrino-$^{4}$He reactions are also investigated by using
the WBP and the SPSDMK Hamiltonians. Calculated cross 
sections are enhanced compared with previous calculations
of ref. \cite{WHHH}. 
A possible consequence of the enhancement of the $\nu$-$^{4}$He
and $\nu$-$^{12}$C reaction cross sections on the abundances 
of light elements is discussed. The production yields of $^{7}$Li 
and $^{11}$B are found to be enhanced during
supernova explosions.

\section*{Acknowledgements}
This work has been supported in part by Grants-in-Aid for Scientific
Research (14540271, 17540275, 18540290) and for Specially Promoted
Research (13002001) of the Ministry of Education, Culture, Sports, 
Science and Technology of Japan, and Mitsubishi Foundation.  
The authors would like to thank Professor A. Gelberg for the careful
reading of the manuscript.



\begin{thebibliography}{99}
\bibitem{SFO03}
T. Suzuki, R. Fujimoto, and T. Otsuka, Phys. Rev. C {\bf 67}, 044302
(2003). 
\bibitem{OUB}
T. Otsuka, R. Fujimoto, Y. Utsuno, B. A. Brown, M. Honma, and
T. Mizusaki, Phys. Rev. Lett. {\bf 87}, 082502 (2001).
\bibitem{OSFGA}
T. Otsuka, T. Suzuki, R. Fujimoto, H. Grawe, and Y. Akaishi, 
Phys. Rev. Lett. {\bf 95}, 232502 (2005). 
\bibitem{VACSG}
C. Volpe, N. Auerbach, G. Col$\grave{o}$, T. Suzuki, and N. Van Giai,
Phys. Rev. C {\bf 62}, 015501 (2000).
\bibitem{HT}
A. C. Hayes and I. S. Towner, Phys. Rev. C {\bf 61}, 044603 (2000). 
\bibitem{WHHH}
S. E. Woosley, D. H. Hartmann, R. D. Hoffman, and W. C. Haxton,
Astrophys. J. {\bf 356}, 272 (1990). 
\bibitem{CK}
S. Cohen and D. Kurath, Nucl. Phys. {\bf 73}, 1 (1965).  
\bibitem{Kir}
M. W. Kirson, Phys. Lett. B {\bf 47}, 110 (1973);\\
I. Kakkar and Y. R. Waghmare, Phys. Rev. C {\bf 2}, 1191 (1970);\\ 
K. Klingenbeck, W. Kn$\ddot{u}$pfer, M. G. Huber, and P. W. M. 
Glaudemans, Phys. Rev. {\bf C15}, 1483 (1977). 
\bibitem{Elliot}
J. P. Elliott, A. D. Jackson, H. A. Mavromatis, E. A. Sanderson, 
and B. Singh, Nucl. Phys. {\bf A121}, 241 (1968). 
\bibitem{PIRHO}
F. Osterfeld, Rev. Mod. Phys. {\bf 64}, 491 (1992);\\
S. -O. B$\ddot{a}$ckman, G. E. Brown, and J. A. Niskanen, Phys. Rep.
{\bf 124}, 1 (1985). 
\bibitem{M3Y}
G. Bertsch, J. Borysowicz, H. McManus, and W. G. Love, 
Nucl. Phys. {\bf A284}, 399 (1977).
\bibitem{Wilk}
R. E. McDonald, J. A. Becker, R. A. Chalmers, and D. H. Wilkinson,
Phys. Rev. C {\bf 10}, 333 (1974). 
\bibitem{DW}
J. D. Walecka, in {\it Muon Physics}, edited by V. H. Highes 
and C. S, Wu (Academic, New York, 1975), Vol. II;\\
J. S. O'Connell, T. W. Donnelly, and J. D. Walecka, Phys. Rev. C
{\bf 6}, 719 (1972);\\
T. W. Donnelly and J. D. Walecka, Nucl. Phys. {\bf A274}, 368 (1976);\\
T. W. Donnelyy and W. C. Haxton, Atomic Data Nucl. Data Tables 
{\bf 23}, 103 (1979). 
\bibitem{Fermi}
D. H. Wilkinson and B. E. F. Macefield, Nucl. Phys. {\bf A232}, 
58 (1974). ¡¡
\bibitem{OXB}
OXBASH, The Oxford, Buenos-Aires, Michigan State, Shell Model Program,
B. A. Brown, A. Etchegoyen, and W. D. M. Rae, MSU Cyclotron Laboratory 
Report No. 524, 1986. 
\bibitem{MK}
D. J. Millener and D. Kurath, Nucl. Phys. {\bf A255}, 315 (1975). 
\bibitem{LSND}
LSND Collaborations, L. B. Auerbach et al., Phys. Rev. C {\bf 64},
065501 (2001). 
\bibitem{KARMEN}
KARMEN Collaboration, B. E. Bodmann et al., Phys. Lett. B {\bf 332}, 
251 (1994).
\bibitem{Navr}
A. C. Hayes, P. Navr$\acute{a}$til, and J. P. Vary, 
Phys. Rev. Lett. {\bf 91}, 12502 (2003). 
\bibitem{Lang}
E. Kolb, K. Langanke, and P. Vogel, Nucl. Phys. {\bf A652}, 91 (1999).
\bibitem{KARM2}
KARMEN Collaboration, B. A. Armbruster et al., Phys. Lett. B {\bf 423}, 15 (1998).
\bibitem{KARM3}
R. Maschuw, Prog. Part. Nucl. Phys. {\bf 40}, 183 (1998).
\bibitem{Drake}
T. E. Drake, E. L. Tomusiak and H. S. Caplan, Nucl. Phys. {\bf A118}, 
138 (1968);\\
A. Yamaguchi, T. Terasawa, K. Nakahara and Y. Torizuka, Phys. Rev. C
{\bf 3},1750 (1971).
\bibitem{Yang}
X. Yang et al., Phys. Rev. C {\bf 48}, 1158 (1993).
\bibitem{Gaard}
C. Gaarde et al., Nucl. Phys. {\bf A422}, 189 (1984). 
\bibitem{Okam}
H. Okamura et al., Phys. Lett. B {\bf 345}, 1 (1995). 
\bibitem{Rapapo}
J. Rapaport, T. Taddeucci, C. Gaarde, C. D. Goodman, C. C. Foster, 
C. A. Goulding, D. Horen, E. Sugarbaker, T. G. Masterson and D. Lind, 
Phys. Rev. C {\bf 24}, 335 (1981).
\bibitem{Comf}
J. R. Comfort, S. M. Austin, P. T. Debevec, G. L. Moake, R. W. Finlay
and W. G. Love, Phys. Rev. C {\bf 21}, 2147 (1980). 
\bibitem{SzSag}
T. Suzuki, H. Sagawa and K. Hagino, Phys. Rev. C {\bf 68}, 014317 (2003).
\bibitem{Pywel}
R. E. Pywell, B. L. Berman, J. G. Woodworth, J. W. Jury, K. G. McNeil
and M. N. Thompson, Phys. Rev. C {\bf 32}, 384 (1985);\\
D. J. Mclean, M. N. Thompson, D. Zubanov, K. G. McNeil, J. W. Jury
and B. L. Berman, {\it ibid.} {\bf 44}, 1137 (1991). 
\bibitem{ASTR}
M. Th. Keil, G. G. Raffelt, and H. -Th. Janka, Astrophys. J. {\bf 590},
971 (2003).   
\bibitem{HF}
W. Hauser and H. Feshbach, Phys. Rev. {\bf 87}, 366 (1952).
\bibitem{WG}
R. L. Walter and P. P. Guss, {\it Proc. Int. Conf. on Nucl. Data for
Basic and Applied Sci.}, Santa Fe, May 13-17, 1985, p.1079 (1986).
\bibitem{AVR94}
V. Avrigeanu and P. E. Hodgson, Phys. Rev. C {\bf 49}, 2136 (1994).
\bibitem{RIPL2}
T. Belgya, O. Bersillon, R. Capote, T. Fukahori, G. Zhigang, S. Goriely,
M. Herman, A. V. Ignatyuk, S. Kailas, A. Koning, P. Oblozhinsky, 
V. Plujko, and P. Young, "{\it Handbook for calculations of nuclear
reaction data: Reference Input Parameter Library}", available online
at http://www-nds.iaea.org/RIPL-2/, IAEA, Vienna (2005).  
\bibitem{WB}
E. K. Warburton and B. A. Brown , Phys. Rev. C {\bf 46}, 923 (1992).
\bibitem{PW}
B. M. Preedom and B. H. Wildenthal, Phys. Rev. C {\bf 6}, 1633 (1972). 
\bibitem{Kuo}
T. T. S. Kuo, Nucl. Phys. {\bf A103}, 71 (1967).
\bibitem{AV8P}
B. S. Pudlinger, V. R. Pandharipande, J. Carlson, S. C. Pieper, and
R. B. Wiringa, Phys. Rev. C {\bf 56}, 1720 (1997).
\bibitem{Gazit}
D. Gazit and N. Barnea, Phys. Rev. C {\bf 70}, 048801 (2004).
\bibitem{Yoshida}
T. Yoshida, T. Kajino, H. Yokomakura, K. Kimura, A. Takamura, and
D. H. Hartmann, Phys. Rev. Lett. {\bf 96}, 091101 (2006).
\bibitem{YKH}
T. Yoshida, T. Kajino, and D. H. Hartmann, Phys. Rev. Lett. {\bf 94}, 231101
(2005). 
\bibitem{Shige}
T. Shigeyama and K. Nomoto, Astrophys. J. {\bf 360}, 242 (1990). 
\bibitem{HW92}
R. D. Hoffman and S. E. Woosley, Neutrino Interaction Cross Sections
and Branching Ratios,
http://www-phys.llnl.gov/Research/RRSN/nu\_csbr/neu\_rate.html, (1992).
\bibitem{YTKS}
T. Yoshida, M. Terasawa, T. Kajino, and K. Sumiyoshi,
Astrophys. J. {\bf 600}, 204 (2004).
\bibitem{Latt}
J. M. Lattimer and M. Prakash,
Astrophys. J. {\bf 550}, 426 (2001). 
\bibitem{YKSC}
T. Yoshida, T. Kajino, T. Suzuki, S. Chiba, T. Otsuka,
A. Takamura, K. Kimura, H. Yokomakura, and D. H. Hartmann, in preparation.

\end{thebibliography}
\end{document}